\documentclass[a4paper,unpublished]{quantumarticle}
\pdfoutput=1
\usepackage{amsmath,amssymb,amsthm,mathtools,bm}
\usepackage{booktabs}
\usepackage{graphicx}
\usepackage{array}
\usepackage{enumitem}
\usepackage{microtype}
\usepackage{xcolor}
\usepackage[colorlinks,bookmarks=false,citecolor=blue,linkcolor=violet,urlcolor=blue]{hyperref}
\usepackage{cleveref}
\usepackage[numbers,sort&compress]{natbib}

\newcommand{\Tr}{\operatorname{Tr}}

\newcommand{\dd}{\mathrm d}
\newcommand{\ii}{\mathrm i}
\newcommand{\e}{\mathrm e}
\newcommand{\HS}{\mathrm{HS}}
\newcommand{\vecop}[1]{|#1\rangle\!\rangle}
\newcommand{\brvec}[1]{\langle\!\langle #1|}

\title{The Pauli Probability Spectrum Carries the Pure-State Quantum Fisher Metric}
\author{E. A. Ramirez Trino}
\affiliation{Instituto de F\'isica, Universidade Federal Fluminense, Av.~Gal.~Milton Tavares de Souza s/n, Gragoat\'a, 24210-346, Niter\'oi, RJ, Brazil}

\author{M. A. Rajabpour}
\affiliation{Instituto de F\'isica, Universidade Federal Fluminense, Av.~Gal.~Milton Tavares de Souza s/n, Gragoat\'a, 24210-346, Niter\'oi, RJ, Brazil}
\affiliation{Theoretical Physics III, Center for Electronic Correlations and Magnetism, Institute of Physics, University of Augsburg, D-86135 Augsburg, Germany}

\begin{document}
\maketitle

\begin{abstract}
How a quantum state responds to small parameter changes underlies state distinguishability, many-body response, and metrological sensitivity. The Pauli probability spectrum provides a natural description of many-body quantum states, but it is not evident how much of this response survives in the corresponding classical distribution. Here we show that, for pure states, the complete labeled Pauli distribution retains the full local metric of the pure-state manifold, reproducing the quantum Fisher metric up to a fixed normalization. This correspondence has a direct measurement realization: preparing a state together with its complex conjugate and performing pairwise Bell measurements between corresponding degrees of freedom converts the quantum response into an ordinary classical distribution without losing any of the sensitivity available in the two inputs. The readout is fixed, factorizes into local pairwise measurements even for strongly entangled many-body states, and preserves the response along all parameter directions simultaneously. We trace this correspondence to the reality of the underlying operator amplitudes and show that the same mechanism extends beyond qubits. When the correspondence fails, we identify explicitly where the missing information resides: part of the response can remain hidden in phases for identical-copy and complex-coordinate readouts, while mixed states develop transverse operator-space motion that is invisible to the Bell probabilities. These results provide a direct link between Pauli/Bell statistics, many-body response, nonstabilizerness, and quantum geometry.
\end{abstract}
\tableofcontents

\section{Introduction}
\label{sec:intro}

Nonstabilizerness has emerged as a useful lens on quantum many-body states, capturing aspects of complexity and critical structure that are not characterized by entanglement alone
\cite{CaoSwingle2021,LeoneOlivieroHamma2022SRE,LiuWinter2022,FrauEtAl2025Disentangling,FalcaoEtAl2025Gauge}.
A central role is played by stabilizer R\'enyi entropies (SREs), which quantify nonstabilizerness through R\'enyi moments of the squared Pauli expectation values
\cite{LeoneOlivieroHamma2022SRE,HaugPiroli2023Monotones}.
This compression is precisely what makes an SRE useful: an exponentially large collection of Pauli weights is reduced to a scalar quantity. At the same time, the Pauli-string labels are discarded. Information contained in how different Pauli sectors respond when the state is varied may therefore be invisible after the compression.

The distinction between a scalar entropy and the probability structure from which it is formed is familiar in many-body physics. Shannon and R\'enyi statistics of critical wave functions encode universal information
\cite{StephanEtAl2009Shannon,StephanMisguichPasquier2011,AlcarazRajabpour2013,AlcarazRajabpour2014,TarighiKhassehNajafiRajabpour2022},
and exact correspondences between stabilizer and Shannon--R\'enyi quantities have recently connected these descriptions directly
\cite{Rajabpour2026SSE}.
Stabilizer-entropic observables further display universal finite-temperature and critical-to-massive crossover structures
\cite{KhassehRajabpour2026FiniteT,KhassehRamirezRajabpour2026Crossovers,KhassehRajabpour2026Fugacity}.
More resolved descriptions reveal additional information: Pauli-spectrum analyses can distinguish structures that are not resolved by a single stabilizer entropy
\cite{OlivieroLeoneHamma2022TFI,Tarabunga2024Critical,TurkeshiEtAlPauliSpectrum,HallamSmithPapic2026,LiChang2026},
while perfect Pauli sampling, Pauli-Markov methods, and matrix-product-state representations in the Pauli basis provide direct access to this finer statistical structure
\cite{LamiCollura2023,TarabungaEtAl2023PauliMarkov,TarabungaEtAl2024PauliMPS,FrauEtAl2024}.
These developments motivate treating the complete labeled Pauli distribution as an information-bearing object in its own right, rather than only as an intermediate object from which a scalar entropy is formed.

The same distribution has also acquired an operational meaning. Bell sampling was introduced in quantum learning of stabilizer states
\cite{Montanaro2017BellSampling}
and has since been developed as a multicopy measurement carrying detailed information about Pauli observables
\cite{Hangleiter2024BellSampling}.
Related protocols provide efficient access to stabilizer entropies through Bell measurements
\cite{HaugLeeKim2024StabilizerEntropies},
and the SRE of critical many-body states can be interpreted as a participation entropy of Bell-measurement probabilities in a doubled Hilbert space
\cite{HoshinoEtAl2026}.
Thus the same underlying data admit two natural levels of description: one may compress the probabilities into R\'enyi moments, or retain their labels and study the full classical distribution.

What information is gained by retaining those labels? A natural way to make this question precise is through information geometry. A smoothly varying classical probability distribution carries a Fisher--Rao geometry, while a family of quantum states has an intrinsic local geometry encoded in the quantum Fisher information matrix (QFIM)
\cite{Helstrom1976,Holevo1982,BraunsteinCaves1994,Liu2019QFIM}.
These two geometries are generally different: fixing a measurement usually discards part of the quantum distinguishability, and in multiparameter estimation measurements that are optimal for different parameter directions need not be mutually compatible
\cite{Matsumoto2002,Ragy2016,Albarelli2019}.
Fisher-symmetric constructions provide balanced access to pure-state parameters within the corresponding information bounds
\cite{LiEtAl2016FisherSymmetric,ZhuHayashi2018UniversallyFisherSymmetric},
whereas real quantum models and antiunitary symmetries identify special situations in which several parameter directions can be accessed compatibly by measurements that do not depend on the unknown target values
\cite{MiyazakiMatsumoto2022,Wang2024AntiunitaryQCRB}.

A connection between this geometric problem and nonstabilizerness has only recently begun to emerge. Stabilizer structure has been related to metrological sensitivity in several settings
\cite{HernandezYanesEtAl2026Metrology,LiraSolanillaEtAl2026ExtensiveQFI},
and at stabilizer states the local curvature of the second SRE under a Hermitian deformation is fixed by the QFI
\cite{JhaGadge2026}.
These results concern selected functions or special points of the Pauli distribution. They leave open a more basic question: \textit{whether the full quantum response is already encoded in the motion of the complete labeled Pauli spectrum, before any R\'enyi compression is performed.}

In this work we answer this question for pure states. We show that the complete labeled Pauli distribution retains the entire local quantum metric, despite losing discrete information about the state at finite separation. The correspondence is therefore geometric rather than tomographic: locally, the classical redistribution of Pauli weights contains all directions in which a pure state becomes distinguishable. From this perspective, an SRE is a compressed probe of a richer geometric object, and its response captures only part of the information carried by the underlying labeled distribution.

We further show that this geometric information has a direct measurement realization. For a state accompanied by its complex-conjugate partner, the relevant operator coordinates become real amplitudes in a doubled Hilbert space, and the Pauli distribution is obtained from pairwise Bell measurements between corresponding sites. This connects the construction to the broader role of real representations and antiunitary symmetry in multiparameter estimation \cite{MiyazakiMatsumoto2022,Wang2024AntiunitaryQCRB}, while retaining a particularly simple many-body architecture: the measurement factorizes site by site even when the state inside each register is strongly entangled. The conjugated state is a physical resource rather than a formal replacement for a second identical copy, and the distinction between the two preparations will be important below.

The operator-space viewpoint also reveals which ingredients of the construction are essential and where the correspondence ceases to hold. Identical replicas need not retain the same multiparameter information because state dependence can remain hidden in phases of the measurement amplitudes, consistently with the information constraints governing collective measurements on identical copies
\cite{GillMassar2000,ZhuHayashi2018UniversallyFisherSymmetric}. Conversely, the Pauli group itself is not fundamental: the same mechanism extends to complete Hermitian operator coordinates in arbitrary finite dimension. Mixed states provide a different boundary, where normalized squared operator coordinates and the probabilities produced by the physical doubled-state measurement no longer coincide. These comparisons separate the role of conjugation, reality, and purity in converting quantum-state motion into classical probability motion.

The resulting framework links three descriptions that are usually treated separately: Pauli-resolved many-body statistics, stabilizer-based measures of nonstabilizerness, and quantum information geometry. The main results are threefold. First, the complete labeled Pauli distribution of a pure state carries the entire local quantum metric, whereas a scalar stabilizer R\'enyi entropy retains only a compressed response of the underlying label-resolved probability flow. Second, for a state together with its complex conjugate, this metric is transferred to an ordinary classical Bell distribution through real operator amplitudes; the corresponding readout factorizes site by site for many-body states and the same mechanism extends to Hermitian operator coordinates beyond qubits. Third, when the conditions behind this correspondence are relaxed, the missing information can be located explicitly: parameter dependence can remain hidden in phases when the relevant amplitudes are complex, including for identical-copy Bell readouts, while mixed states develop transverse operator-space motion that is invisible to the Bell probabilities.

The same labeled record provides a direct route to geometric and statistical inference without an intermediate reconstruction of the quantum state. Pauli measurements are already being used on quantum processors to extract universal properties of critical many-body states \cite{KoyluogluEtAl2026CentralCharge}. Our results determine what information the ideal labeled Pauli/Bell record contains and which physical assumptions allow a fixed measurement to preserve it; they do not by themselves determine the finite-sample cost of extracting that information from the exponentially large outcome space. We return to this distinction in Sec.~\ref{sec:operational}.


\section{Quantum Fisher geometry of the Pauli spectrum}
\label{sec:pure-fisher}

Let $\rho=|\psi\rangle\langle\psi|$ be a pure $N$-qubit state, with $d=2^N$. For each Pauli string $P\in\mathcal P_N=\{I,X,Y,Z\}^{\otimes N}$, the coefficient $a_P=\Tr(P\rho)=\langle\psi|P|\psi\rangle$ is real. The squared Pauli coefficients define the labeled distribution
\begin{equation}
\Xi_\rho(P)=\frac{1}{d}[\Tr(P\rho)]^2=\frac{a_P^2}{d}.
\label{eq:XiDefinition}
\end{equation}
Using $\Tr(PP')=d\,\delta_{P,P'}$, the state and its purity can be written as
\begin{equation}
\begin{split}
\rho=\frac{1}{d}\sum_{P\in\mathcal P_N}a_P P,\quad
\Tr(\rho^{2})=\frac{1}{d}\sum_{P\in\mathcal P_N}a_P^2 .
\end{split}
\label{eq:pauliexpansion}
\end{equation}
For a pure state $\Tr\rho^{2}=1$, so $\Xi_\rho$ is a normalized probability distribution over Pauli strings.

We use $\Xi_\rho(P)$ for the probability assigned to a given Pauli string and $\Xi_\rho\equiv\{\Xi_\rho(P)\}_{P\in\mathcal P_N}$ for the complete labeled distribution. Retaining the labels will be essential when comparing its motion with scalar functions of the Pauli weights. A compact summary of the notation used throughout the manuscript is given in Appendix~\ref{app:notation}.

To study the response of the Pauli distribution, we make the parameter dependence of the state explicit and write \(\rho\equiv\rho_{\boldsymbol\theta}=|\psi_{\boldsymbol\theta}\rangle\langle\psi_{\boldsymbol\theta}|\), with \(\boldsymbol\theta=(\theta^1,\ldots,\theta^m)\) and $\partial_\mu\equiv\partial/\partial\theta^\mu$. The components of $\boldsymbol\theta$ may represent, for example, couplings or external fields in a many-body Hamiltonian. For a smooth discrete distribution $p_\ell(\boldsymbol\theta)$, where $\ell$ labels its possible outcomes, the classical Fisher matrix is \(F_{\mu\nu}[p]=\sum_{\ell:p_\ell>0}\partial_\mu p_\ell\,\partial_\nu p_\ell/p_\ell\). 
It quantifies the local distinguishability of nearby probability distributions \cite{Fisher1922,FisherRaoInfoGeometry}. The corresponding intrinsic distinguishability of pure quantum states is encoded in the QFIM, \(F^{\rm Q}_{\mu\nu}[\rho]=4\,\Re\!\left[\langle\partial_\mu\psi|\partial_\nu\psi\rangle- \langle\partial_\mu\psi|\psi\rangle\langle\psi|\partial_\nu\psi\rangle\right]\)~\cite{BraunsteinCaves1994,Liu2019QFIM}. We first consider regular points; the continuous extension through vanishing Pauli probabilities is discussed in Appendix~\ref{app:zeros}.

For the Pauli distribution, the relation between the classical and quantum responses follows directly from its operator coordinates. Since $\Xi_\rho(P)=a_P^2/d$, \(F_{\mu\nu}[\Xi_\rho]=4\sum_P(\partial_\mu a_P)(\partial_\nu a_P)/d\). Orthogonality of the Pauli basis converts this sum into the Hilbert--Schmidt length of the state variation, \(F_{\mu\nu}[\Xi_\rho]=4\,\Tr[(\partial_\mu\rho)(\partial_\nu\rho)]\). For a pure projector we have, \(F^{\rm Q}_{\mu\nu}[\rho]=2\,\Tr[(\partial_\mu\rho)(\partial_\nu\rho)]\), and therefore
\begin{equation}
F_{\mu\nu}[\Xi_\rho]=2F^{\rm Q}_{\mu\nu}[\rho].
\label{eq:PauliFisherTheorem}
\end{equation}
Equation~\eqref{eq:PauliFisherTheorem} is the central pure-state identity of this work. The complete labeled Pauli distribution is locally metric-complete: its classical Fisher geometry retains the full QFIM of the state, up to the fixed factor of two, simultaneously for all parameter directions.

\subsection{Local quantum geometry and finite-distance sign loss}
\label{sec:local-global}

The identity has an immediate geometric interpretation. The quantum geometric tensor is \(Q_{\mu\nu}=\langle\partial_\mu\psi|(I-\rho)|\partial_\nu\psi\rangle\), and \(g_{\mu\nu}=\Re Q_{\mu\nu}\), with $F^{\rm Q}_{\mu\nu}=4g_{\mu\nu}$. Since the quantum metric governs the leading distinguishability of nearby states and underlies fidelity susceptibility and geometric response in many-body systems \cite{VenutiZanardi2007,KolodrubetzEtAl2017}, Eq.~\eqref{eq:PauliFisherTheorem} shows that the same local geometry is encoded in the motion of the Pauli distribution. A differential-geometric formulation is given in Appendix~\ref{app:finitegeometry}.

Local metric completeness does not imply global reconstruction of the state: the map $a_P\mapsto a_P^2$ removes the signs of the Pauli coordinates. For two pure states $\rho=|\psi\rangle\langle\psi|$ and $\sigma=|\varphi\rangle\langle\varphi|$, define their Pauli-distribution overlap as $B_{\rm P}(\rho,\sigma)=\sum_P\sqrt{\Xi_\rho(P)\Xi_\sigma(P)}$. Writing $a_P=\Tr(P\rho)$ and $b_P=\Tr(P\sigma)$ gives
\begin{equation}
B_{\rm P}(\rho,\sigma)=\frac{1}{d}\sum_P|a_Pb_P|\geq\Tr(\rho\sigma)=|\langle\psi|\varphi\rangle|^2 .
\label{eq:BhattacharyyaGlobal}
\end{equation}
The inequality isolates the information lost by squaring: the Pauli distribution retains the coordinate magnitudes but not their relative signs. It becomes an equality whenever the nonzero Pauli coordinates of the two states have the same sign structure. The derivation and the behavior across zero crossings are given in Appendix~\ref{app:finitegeometry}.

For neighboring states at a regular point this discrete ambiguity is absent locally, and
\begin{equation}
1-B_{\rm P}(\rho_{\boldsymbol\theta},\rho_{\boldsymbol\theta+\dd\boldsymbol\theta})=\frac{1}{4}F^{\rm Q}_{\mu\nu}[\rho_{\boldsymbol\theta}]\,\dd\theta^\mu\dd\theta^\nu+O(\|\dd\boldsymbol\theta\|^3).
\label{eq:HellingerLocal}
\end{equation}
Thus the curvature of the overlap between neighboring Pauli distributions directly reproduces the quantum metric. For a single parameter $\lambda$, this relation reads $F^{\rm Q}_{\lambda\lambda}=4\lim_{\epsilon\to0}[1-B_{\rm P}(\rho_\lambda,\rho_{\lambda+\epsilon})]/\epsilon^2$. This is the finite-distance counterpart of Eq.~\eqref{eq:PauliFisherTheorem}; the expansion yielding the factor $1/4$ is given in Appendix~\ref{app:finitegeometry}.

\subsection{What survives under R\'enyi compression}
\label{sec:renyi-compression}

The Fisher identity concerns the complete labeled distribution. We now ask how much of this geometric information survives when $\Xi_\rho$ is compressed to a single R\'enyi entropy,
\begin{equation}
H_\alpha(\Xi_\rho)=\frac{1}{1-\alpha}\log\!\sum_P\Xi_\rho(P)^\alpha,
\end{equation}
where \(\alpha>0\), and \(\alpha\neq1\). The stabilizer R\'enyi entropy convention of Ref.~\cite{LeoneOlivieroHamma2022SRE} is $M_\alpha=H_\alpha-\log d$. Since this shift is independent of $\boldsymbol\theta$, the two quantities have identical parameter derivatives.

This compression removes the label-resolved structure that carries the full Fisher geometry. While the Fisher matrix is sensitive to the motion of every Pauli probability, $H_\alpha$ retains only one weighted moment of them; compensating changes among different Pauli sectors can therefore leave the entropy stationary.

For a trajectory $\boldsymbol\theta=\boldsymbol\theta(\lambda)$, let $v^\mu=\dd\theta^\mu/\dd\lambda$, so that $\partial_vH_\alpha=v^\mu\partial_\mu H_\alpha$ and $F^{\rm Q}_v[\rho]=v^\mu F^{\rm Q}_{\mu\nu}[\rho]v^\nu$. This simply selects the physical deformation along which the response is evaluated.

For $\alpha>1/2$ and $\alpha\neq1$, the R\'enyi response satisfies
\begin{equation}
F^{\rm Q}_v[\rho]\geq\frac{(1-\alpha)^2}{2\alpha^2}e^{-2(1-\alpha)\left(H_{2\alpha-1}-H_\alpha\right)}|\partial_v H_\alpha|^2 .
\label{eq:SREgradientBound}
\end{equation}
The same bound holds for $M_\alpha$, because the constant shift $-\log d$ cancels from both the derivative and the difference $M_{2\alpha-1}-M_\alpha$. Equation~\eqref{eq:SREgradientBound} makes the asymmetry between the full distribution and its R\'enyi compression explicit: a nonzero entropy response imposes a lower bound on the QFI along the same deformation, whereas a finite QFI need not produce a nonzero $\partial_vH_\alpha$. The underlying labeled probabilities may continue to move while their contributions to the R\'enyi moment cancel.

The derivation of Eq.~\eqref{eq:SREgradientBound}, including the weighted distribution and logarithmic derivatives entering the Cauchy--Schwarz step, is given in Appendix~\ref{app:renyi-response}. The Shannon limit and the treatment of vanishing probabilities are discussed in Appendices~\ref{app:shannon-bound} and \ref{app:zeros}, respectively.

The loss under R\'enyi compression is therefore one-sided. A nonzero R\'enyi response certifies underlying state motion and constrains its QFI, but a stationary R\'enyi entropy does not imply a stationary Pauli distribution. Even the complete set of R\'enyi entropies at a fixed parameter value determines only the unordered Pauli weights, not which labeled sectors carry the motion along a trajectory.

\section{Bell sampling of the Pauli spectrum}
\label{sec:bell}

We now turn Eq.~\eqref{eq:PauliFisherTheorem} into an operational statement. The complete Pauli distribution is directly sampled by a fixed pairwise Bell measurement on a pure state and its complex conjugate, closely related to Bell-sampling protocols for stabilizer R\'enyi entropies \cite{Montanaro2017BellSampling,HaugLeeKim2024StabilizerEntropies}.

For a single qubit, the same state can be written in Bloch form as $\rho=\frac12(I+xX+yY+zZ)$, with $x^2+y^2+z^2=1$. Prepare $|\psi\rangle\otimes|\psi^*\rangle\equiv|\psi\rangle|\psi^*\rangle$, where conjugation is taken in the computational basis defining the Pauli operators. Labeling the Bell outcomes by $I,X,Y,Z$, their probabilities are \(q^{\rm conj}(I,X,Y,Z)=\frac12(1,x^2,y^2,z^2)=\Xi_\rho\). Thus the squared Pauli expectation values appear directly as measurement probabilities. The Bell-state conventions, projectors, normalization, and an explicit one-qubit check are given in Appendix~\ref{app:qubitchecks}.

For an interacting $N$-qubit state, prepare two registers $A$ and $B$ in $|\psi\rangle_A$ and $|\psi^*\rangle_B$ and perform Bell measurements between corresponding sites. The local outcomes define a Pauli string $P=P_1\otimes\cdots\otimes P_N$. Using the standard operator--state correspondence, with column vectorization $\vecop{O}=\sum_{ij}O_{ij}|i\rangle_A|j\rangle_B$ \cite{Jamiolkowski1972,Choi1975}, a pure projector satisfies $\vecop{\rho}=|\psi\rangle_A|\psi^*\rangle_B$, while the normalized vectors $\vecop{P}/\sqrt d$ form the Pauli-labeled Bell basis. Hence
\begin{equation}
\begin{aligned}
q^{\rm conj}(P)
&=
\frac1d
\left|
\langle\!\langle P|\rho\rangle\!\rangle
\right|^2
\\
&=
\frac1d
[\Tr(P\rho)]^2
=
\Xi_\rho(P).
\end{aligned}
\label{eq:BellXi}
\end{equation}
The projectors and normalization conventions underlying Eq.~\eqref{eq:BellXi} are given in Appendix~\ref{app:qubitchecks}.

Because vectorization preserves the tensor-product structure, $\vecop{P}/\sqrt{2^N}=\bigotimes_{j=1}^{N}\vecop{P_j}/\sqrt2$. The global readout therefore factorizes into $N$ Bell measurements between corresponding sites, even when each register contains an arbitrarily entangled many-body state.

Combining Eq.~\eqref{eq:BellXi} with Eq.~\eqref{eq:PauliFisherTheorem} gives the measurement-level consequence. Additivity of the QFIM and invariance of its metric part under complex conjugation imply $F^{\rm Q}[\rho\otimes\rho^*]=F^{\rm Q}[\rho]+F^{\rm Q}[\rho^*]=2F^{\rm Q}[\rho]$. Therefore
\begin{equation}
F_{\rm Bell}=2F^{\rm Q}[\rho],
\label{eq:BellOptimal}
\end{equation}
where \(F_{\rm Bell}\equiv F[q^{\rm conj}]\). Equation~\eqref{eq:BellOptimal} is the central operational consequence of the Pauli--Fisher identity: a single fixed pairwise Bell readout extracts the complete local quantum sensitivity available in the conjugate pair, simultaneously for all parameter directions and without adaptation to the working point. The factor of two reflects the two parameter-dependent inputs and is not an enhancement of Fisher information per copy.

Complex conjugation also clarifies why the same fixed readout can access all parameter directions. The quantum geometric tensor introduced in Sec.~\ref{sec:local-global} contains a real symmetric part, which gives the quantum metric, and an imaginary antisymmetric part. Complex conjugation leaves the metric part unchanged while reversing the sign of the imaginary part. Hence, for the conjugate pair, \(Q_{\mu\nu}^{\rm conj}=Q_{\mu\nu}+Q_{\mu\nu}^*=2g_{\mu\nu}\). The metric contributions therefore add, while the antisymmetric contributions cancel. This is consistent with the role of antiunitary symmetry in compatible multiparameter estimation \cite{MiyazakiMatsumoto2022}; Eq.~\eqref{eq:BellXi} realizes the resulting metric through a fixed local Bell readout.

The preparation requirement remains essential: the second register must realize the family $|\psi_{\boldsymbol\theta}^*\rangle$ in the fixed reference basis defining the operator coordinates; no universal operation that complex-conjugates an arbitrary unknown state is assumed. Since $|\psi^*\rangle$ is generally distinct from $|\psi\rangle$, the next question is what geometry remains accessible when only identical copies are available.

\section{Fisher information from identical copies}
\label{sec:identical}

The fixed Bell readout of Sec.~\ref{sec:bell} retains the complete QFIM of a state and its complex conjugate. Replacing the conjugated register by an identical copy changes this conclusion: in a fixed measurement basis, parameter dependence stored in relative phases need not appear in the outcome probabilities.

\subsection{Amplitude, phase, and the Fisher-information budget}

To isolate this distinction, expand the pure-state family in a fixed orthonormal basis as $|\psi_{\boldsymbol\theta}\rangle=\sum_a c_a|e_a\rangle$, with $c_a=r_a e^{\ii\phi_a}$ and $p_a=r_a^2$. A projective measurement in this basis sees only the probabilities $p_a$, whose Fisher matrix is $F_{\mu\nu}[p]=4\sum_a(\partial_\mu r_a)(\partial_\nu r_a)$. Parameter dependence carried only by the relative phases $\phi_a$ is therefore invisible to this measurement. For a pure state, the missing contribution is explicit:
\begin{equation}
F^{\rm Q}_{\mu\nu}[\rho]=F_{\mu\nu}[p]+4\operatorname{Cov}_{p}
(\partial_\mu\phi_a,\partial_\nu\phi_a).
\label{eq:phaseCov}
\end{equation}
Here $\operatorname{Cov}_{p}(A,B)=\langle AB\rangle_p-\langle A\rangle_p\langle B\rangle_p$. The covariance removes common phase motion, leaving only changes of relative phase. Equation~\eqref{eq:phaseCov} is the amplitude--phase decomposition of the pure-state Fisher geometry \cite{Luo2006FisherWavefunctions}; its derivation is given in Appendix~\ref{app:phase}.

This distinction directly explains the role of conjugation in Sec.~\ref{sec:bell}. For $|\psi\rangle|\psi^*\rangle$, the Pauli-labeled Bell amplitudes are $\Tr(P\rho)/\sqrt d$ and are real, so their parameter dependence is fully visible in the outcome probabilities. For $|\psi\rangle|\psi\rangle$, the corresponding Bell amplitudes are generally complex, and part of the state motion can remain in their phases. The operator-space form of this distinction is given in Appendix~\ref{app:phase}.

The same limitation has a global multiparameter form. Let $F^{(n)}$ be the classical Fisher matrix of a measurement on $n$ identical copies and $F^{\rm Q}[\rho]$ the one-copy QFIM. The dimensionless quantity $\Tr[(F^{\rm Q}[\rho])^{-1}F^{(n)}]$ measures the total Fisher information recovered relative to the one-copy quantum metric; in local coordinates with $F^{\rm Q}[\rho]=I$, it is the sum over the $2(d-1)$ real tangent directions of a pure state in dimension $d$. The Gill--Massar bound gives \cite{GillMassar2000}
\begin{equation}
\Tr\!\left[\left(F^{\rm Q}[\rho]\right)^{-1}F^{(n)}\right]
\leq n(d-1).
\label{eq:GM}
\end{equation}
By comparison, the complete $n$-copy QFIM $nF^{\rm Q}[\rho]$ gives $2n(d-1)$ in the same normalized trace. Thus a single measurement on identical copies cannot reproduce the complete $n$-copy QFIM simultaneously over the full pure-state manifold. This is a constraint on the total multiparameter information and does not preclude optimal sensitivity along individual directions.

Different measurements distribute this allowed information differently. Fisher-symmetric constructions distribute it uniformly among local directions \cite{ZhuHayashi2018UniversallyFisherSymmetric}; for two identical copies their optimal pure-state construction gives $F^{(2)}=F^{\rm Q}[\rho]$ in the normalization used here. Appendix~\ref{app:GMproof} gives a direct derivation of Eq.~\eqref{eq:GM}. We refer to its right-hand side as the \emph{Fisher-information budget}: it constrains the total information available to a single identical-copy measurement without fixing its distribution among parameter directions.

\subsection{Bell-like measurements on multiple identical copies}
\label{sec:relative-weyl}

We now ask whether increasing the number of identical copies can make a Bell-like fixed readout approach the conjugate-pair performance of Sec.~\ref{sec:bell}. For even $n$, consider the commuting observables $X^{\otimes n}$ and $Z_jZ_n$, $j=1,\ldots,n-1$. Their common eigenstates are \(|[\mathbf b],\pm\rangle=\frac{|\mathbf b\rangle\pm|\bar{\mathbf b}\rangle}{\sqrt2}\), \([\mathbf b]=\{\mathbf b,\bar{\mathbf b}\}\), where $\mathbf b\in\{0,1\}^n$ and $\bar{\mathbf b}$ is its bitwise complement. We call this the relative-Weyl readout. For $n=2$ it reduces to the ordinary Bell basis. Its algebraic construction, completeness, replica-permutation symmetry, and even--odd structure are derived in Appendix~\ref{app:weyl}.

For a single qubit, parameterize the state as \(|\psi(\vartheta,\varphi)\rangle=\cos\frac{\vartheta}{2}|0\rangle
+e^{\ii\varphi}\sin\frac{\vartheta}{2}|1\rangle\) where $\vartheta$ changes the basis populations and $\varphi$ their relative phase. The one-copy and $n$-copy QFIMs are
\[
F^{\rm Q,(1)}
=
\begin{pmatrix}
1 & 0\\
0 & \sin^2\vartheta
\end{pmatrix},
\qquad
F^{\rm Q,(n)}
=
nF^{\rm Q,(1)}.
\]
Let $F^{(n)}$ denote the Fisher matrix of the relative-Weyl readout on $n$ identical copies. Appendix~\ref{app:weylFI} gives
\begin{subequations}
\begin{align}\label{eq:traceidentity}
F_{\vartheta\vartheta}^{(n)}+\frac{F_{\varphi\varphi}^{(n)}}{\sin^2\vartheta}=n,\\
F_{\varphi\varphi}^{(n)}\leq n\sin^n\vartheta \label{eq:phasebound}
\end{align}
\end{subequations}
Equation~\eqref{eq:traceidentity} shows that the relative-Weyl readout exhausts the qubit Fisher-information budget, while Eq.~\eqref{eq:phasebound} determines how that budget is distributed. In particular, $F_{\varphi\varphi}^{(n)}/[n\sin^2\vartheta]\leq\sin^{\,n-2}\vartheta$, so for any fixed $\vartheta\neq\pi/2$ the fraction of available phase information decreases exponentially with $n$. Since the total normalized Fisher information remains fixed by Eq.~\eqref{eq:traceidentity}, the readout becomes increasingly concentrated along the population-changing direction. Positivity also suppresses the normalized off-diagonal component. The finite-$n$ expressions and large-$n$ limit are given in Appendix~\ref{app:weylFI}.

Near the equator, $\sin\vartheta\simeq1$, and the phase-sensitive region narrows only on the scale $|\vartheta-\pi/2|=O(n^{-1/2})$, as derived in Appendix~\ref{app:weylFI} and shown in Fig.~\ref{fig:relativeWeyl}.

\begin{figure}[!ht]
\centering
\includegraphics[width=\columnwidth]{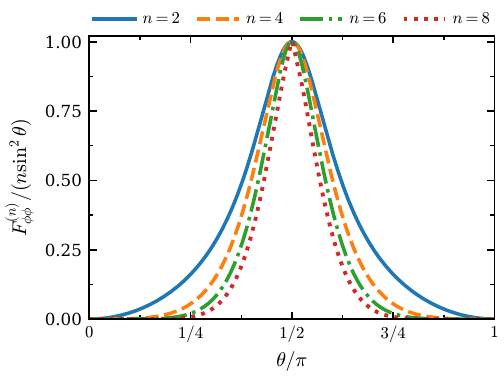}
\caption{\textbf{Identical-copy Fisher redistribution.}
Fraction $F_{\varphi\varphi}^{(n)}/[n\sin^2\vartheta]$ of the available $n$-copy QFI recovered in the phase direction by the relative-Weyl readout for $n=2,4,6,8$ and $\varphi=0.37$. Increasing the number of identical copies confines appreciable phase sensitivity to an increasingly narrow region around $\vartheta=\pi/2$.}
\label{fig:relativeWeyl}
\end{figure}

Increasing the number of identical copies therefore does not drive this Bell-like hierarchy toward the conjugate-pair geometry. The relative-Weyl readout continues to exhaust the qubit Fisher-information budget but becomes increasingly anisotropic, with phase sensitivity suppressed away from the equator. 

\section{Why the Bell construction works: real amplitudes beyond qubits}
\label{sec:frames}

The Bell construction of Sec.~\ref{sec:bell} admits a simple operator-space interpretation that reveals which of its ingredients are essential. The Pauli algebra itself is not: what matters is that the conjugate pair $\vecop{\rho}=|\psi\rangle|\psi^*\rangle$ can be expanded in a complete set of real operator coordinates. Hermitian operator bases provide such coordinates automatically and extend the same fixed-readout mechanism beyond qubits.

\subsection{Real operator coordinates carry the full metric}

Let $\mathcal A=\{A_a\}_{a=1}^{d^2}$ be a Hermitian Hilbert--Schmidt orthonormal basis for operators on a $d$-dimensional Hilbert space,
\(A_a=A_a^\dagger\), \(\Tr(A_aA_b)=\delta_{ab}\).
The density matrix has the expansion $\rho=\sum_a\eta_aA_a$, with
\(\eta_a=\Tr(A_a\rho)\in\mathbb R\).
For a pure state, $\sum_a\eta_a^2=\Tr(\rho^2)=1$, so the squared coordinates define the labeled probability distribution
\(\Xi_{\mathcal A}(a)=\eta_a^2=[\Tr(A_a\rho)]^2\).
For the normalized Pauli basis $A_P=P/\sqrt d$, this reduces to $\Xi_\rho(P)$ of Sec.~\ref{sec:pure-fisher}.

The Fisher identity follows from completeness exactly as in the Pauli case:
\(F_{\mu\nu}[\Xi_{\mathcal A}]
=4\sum_a(\partial_\mu\eta_a)(\partial_\nu\eta_a)
=4\Tr[(\partial_\mu\rho)(\partial_\nu\rho)]\).
Using the pure-state relation
$F^{\rm Q}_{\mu\nu}[\rho]
=2\Tr[(\partial_\mu\rho)(\partial_\nu\rho)]$,
\begin{equation}
F_{\mu\nu}[\Xi_{\mathcal A}]
=
2F^{\rm Q}_{\mu\nu}[\rho].
\label{eq:HSFisher}
\end{equation}
Thus the Pauli--Fisher identity is a consequence of real, complete operator coordinates rather than of the multiplication rules of the Pauli group.

The same observation gives the corresponding measurement directly. Since
\(\vecop{\rho}=|\psi\rangle|\psi^*\rangle\) and
\(\langle\!\langle A_a|\rho\rangle\!\rangle
=\Tr(A_a\rho)=\eta_a\),
the amplitudes of the conjugate pair in the vectorized Hermitian basis are real by construction. A projective measurement in $\{\vecop{A_a}\}$ therefore produces
\(q_{\mathcal A}(a)=|\langle\!\langle A_a|\rho\rangle\!\rangle|^2
=\Xi_{\mathcal A}(a)\), and hence
\begin{equation}
F[q_{\mathcal A}]=F^{\rm Q}[\rho\otimes\rho^*]=2F^{\rm Q}[\rho].
\label{eq:HermitianReadoutOptimal}
\end{equation}
Equation~\eqref{eq:HermitianReadoutOptimal} identifies the Pauli/Bell protocol as the qubit realization of a more general Hermitian-coordinate readout.

This construction places the Pauli/Bell result within the broader theory of real quantum models, antiunitary symmetry, and symmetry-induced optimal measurements \cite{MiyazakiMatsumoto2022,Wang2024AntiunitaryQCRB,LiuShiWuYu2026StateSymmetry}. Such structures are known to permit fixed or target-independent measurements that attain multiparameter quantum limits in suitable state families. Here the additional structure is explicit at the level of the measured probability distribution: vectorization maps the conjugate pair to the Hermitian projector $\rho$, while a Hermitian operator basis provides real amplitudes whose squared values form a classical distribution carrying the complete pure-state metric. This viewpoint also exposes what is lost when the conditions are relaxed: complex operator coordinates can retain parameter dependence in phases, whereas mixed states introduce a distinct transverse component of the operator-space motion. Choosing the Hermitian basis locally further makes the many-body factorization of the readout transparent.

\subsection{Local measurements beyond qubits}

The construction remains local across the two registers in arbitrary finite on-site dimension. Consider
\(\mathcal H=\bigotimes_{j=1}^{N}\mathcal H_j\),
with \(\dim\mathcal H_j=d_j\), and choose a Hermitian Hilbert--Schmidt orthonormal basis
$\mathcal A_j=\{A_{a_j}^{(j)}\}_{a_j=1}^{d_j^2}$
at each site. Their tensor products
\(A_{\boldsymbol a}
=A_{a_1}^{(1)}\otimes\cdots\otimes A_{a_N}^{(N)}\)
form a Hermitian orthonormal basis for the full operator space, and vectorization preserves the product structure,
\(\vecop{A_{\boldsymbol a}}
=\bigotimes_{j=1}^{N}\vecop{A_{a_j}^{(j)}}\).
The global optimal readout therefore factorizes into independent measurements between corresponding local degrees of freedom, even when the state within either register is many-body entangled.

A canonical local Hermitian basis exists in every finite dimension. For a reference basis $\{|i\rangle\}_{i=0}^{d_{\rm loc}-1}$, its vectorized measurement states may be chosen as
\[
|ii\rangle,
\qquad
\frac{|ij\rangle+|ji\rangle}{\sqrt2},
\qquad
\frac{-\ii|ij\rangle+\ii|ji\rangle}{\sqrt2},
\qquad
i<j.
\]
These $d_{\rm loc}^2$ states resolve the diagonal matrix elements and the real and imaginary parts of every coherence. Their associated operators and explicit qutrit and spin-$s$ realizations are given in Appendix~\ref{app:higherdim}. For $d_{\rm loc}=2$, an orthogonal rotation within the diagonal sector recovers the ordinary Bell basis.

Unlike the qubit Pauli realization, the higher-dimensional readout need not consist entirely of maximally entangled vectors: for example, the states $|ii\rangle$ are product states. Nevertheless, the complete measurement still attains Eq.~\eqref{eq:HermitianReadoutOptimal}. This separates Fisher saturation from the entanglement structure of the individual measurement vectors and motivates identifying the operator property that controls saturation.

\subsection{Why Hermiticity matters}

The amplitude--phase distinction of Sec.~\ref{sec:identical} has an exact operator-space counterpart. Let $\mathcal B=\{B_a\}_{a=1}^{d^2}$ be an orthonormal operator basis that is not assumed Hermitian, \(\Tr(B_a^\dagger B_b)=\delta_{ab}\), and, on the regular support $\zeta_a\neq0$, write its generally complex coordinates as \(\zeta_a=\Tr(B_a^\dagger\rho)=u_a e^{\ii\varphi_a}\). For sectors with $\zeta_a=0$, the polar variables are singular, and the metric identities below are understood by continuous extension from nearby regular points, as in Appendix~\ref{app:zeros}. The corresponding vectorized-basis measurement records only \(\Gamma_{\mathcal B}(a)=|\zeta_a|^2=u_a^2\). Completeness then gives the exact deficit
\begin{equation}
2F^{\rm Q}_{\mu\nu}[\rho]
-
F_{\mu\nu}[\Gamma_{\mathcal B}]
=
4\sum_a
u_a^2
(\partial_\mu\varphi_a)
(\partial_\nu\varphi_a).
\label{eq:complexdeficit}
\end{equation}
The right-hand side is positive semidefinite and is the operator-space analogue of Eq.~\eqref{eq:phaseCov}: parameter dependence stored only in coordinate phases contributes to the quantum metric but is absent from the measured probabilities. For a Hermitian basis, $\zeta_a=\Tr(A_a\rho)$ is real, the phase term vanishes, and Eq.~\eqref{eq:HSFisher} is recovered. The derivation and its extension to Hermitian tight frames are given in Appendix~\ref{app:complexframes}.

A generalized Weyl basis provides a useful contrast. Its operators are unitary and its vectorized states are maximally entangled, yet its operator coordinates are generally complex and Eq.~\eqref{eq:complexdeficit} need not vanish. Maximal entanglement of the measurement basis is therefore not sufficient for Fisher saturation.

Within orthogonal operator-basis readouts, Hermiticity is also essentially necessary for universal pure-state saturation. Appendix~\ref{app:frameconverse} shows that state-independent saturation forces each basis operator to be Hermitian up to a fixed phase. The Pauli basis adds two useful properties to this saturation condition: unitarity makes its vectorized states Bell-like, while its tensor-product structure makes the global readout sitewise factorized.

\section{Mixed states and the Bell-information gap}
\label{sec:mixeddistributions}

We now allow $\rho$ to denote a generally mixed density operator; when its parameter dependence must be displayed explicitly we write $\rho_{\boldsymbol\theta}$. Purity was responsible for two coincidences used throughout the preceding sections: the squared Pauli coordinates were already normalized, and the same numbers were the physical Bell probabilities of the conjugate pair. Away from rank one these statements separate, and so do their Fisher geometries.

\subsection{Three distributions that separate away from purity}

Let $R=\Tr(\rho^2)$ denote the purity and $a_P=\Tr(P\rho)$ the real Pauli coefficients. Since $\sum_Pa_P^2=dR$, their normalized squares define
\begin{equation}
\Xi_\rho^{\HS}(P)
=
\frac{a_P^2}{dR}.
\label{eq:XiHSmixed}
\end{equation}
The distribution $\Xi_\rho^{\HS}$ describes the normalized operator $\rho/\sqrt R$ in Hilbert--Schmidt space. It is a legitimate probability distribution, but it is not, in general, the outcome distribution of a Bell measurement on two physical registers.

For the physical conjugate pair $\rho\otimes\rho^*$, the Pauli-labeled Bell readout of Sec.~\ref{sec:bell} instead gives

\begin{equation}
q^{\rm conj}(P)
=
\frac1d\Tr(P\rho P\rho).
\label{eq:qconj}
\end{equation}
For a pure projector, $\rho P\rho=\Tr(P\rho)\rho$, and Eq.~\eqref{eq:qconj} reduces to $q^{\rm conj}(P)=\Xi_\rho(P)$. If the two registers are identically prepared $\rho\otimes\rho$, the same Bell measurement gives
\begin{equation}
q^{\rm id}(P)
=
\frac1d\Tr(P\rho P\rho^\top).
\label{eq:qid}
\end{equation}
Thus the single pure-state Pauli distribution separates into three probability models once purity is lost. Their relations in the relevant limits are summarized in Table~\ref{tab:mixeddistributions}.

\begin{table}[!ht]
\caption{\textbf{Mixed-state probability dictionaries.}
Relations among the normalized squared Pauli coordinates and the two physical Bell distributions. ``Real'' refers to the computational basis defining complex conjugation.}
\label{tab:mixeddistributions}
\centering
\begin{tabular}{@{}ll@{}}
\hline
State & Relation \\ \hline
mixed, complex & generically all distinct \\
mixed, real &
$q^{\rm id}=q^{\rm conj}\neq\Xi_\rho^{\HS}$ \\
pure, complex &
$q^{\rm id}\neq q^{\rm conj}=\Xi_\rho^{\HS}=\Xi_\rho$ \\
pure, real &
$q^{\rm id}=q^{\rm conj}=\Xi_\rho^{\HS}=\Xi_\rho$ \\
\hline
\end{tabular}
\end{table}

The distinction between $\Xi_\rho^{\HS}$ and $q^{\rm conj}$ is already visible in the identity outcome: $q^{\rm conj}(I)=R/d$, whereas $\Xi_\rho^{\HS}(I)=1/(dR)$. Their equality therefore requires $R=1$. Identifying normalized squared Pauli coordinates with physical conjugate-copy Bell probabilities is a genuinely rank-one property.

The distributions are nevertheless related algebraically. In particular, Appendix~\ref{app:mixeddictionary} constructs a signed linear map $\mathsf M$ such that $q^{\rm conj}=\mathsf M q^{\rm id}$. Exact identical-copy probabilities therefore determine the ideal conjugate-copy distribution. This does \emph{not} make the two experiments statistically equivalent. The map contains negative entries and is not a stochastic channel: when applied to empirical frequencies it transforms their sampling fluctuations together with their mean. The resulting estimator reconstructs $q^{\rm conj}$ but does not acquire the multinomial statistics of direct conjugate-copy Bell sampling. Consequently, classical post-processing cannot restore Fisher information that was absent from the identical-copy measurement.

This distinction is already relevant for pure complex states. There $q^{\rm conj}=\Xi_\rho$ carries the complete doubled-state QFIM, while $q^{\rm id}$ can be mapped to the same \emph{ideal} probability vector without carrying the same Fisher information. Only when the state is real in the reference basis do the two physical experiments themselves coincide. In the infinite-sample limit the reconstructed frequencies converge to $q^{\rm conj}$ because the fluctuations vanish, but the finite-resource precision remains governed by the Fisher information of the measurement that was actually performed.

Among these distributions, $q^{\rm conj}$ is therefore the physical continuation of the readout that was optimal in the pure-state problem. We next identify exactly which part of the quantum response becomes invisible to it.

\subsection{Where the Bell information is lost}
\label{sec:mixedgap}

Let $\Omega=\rho\otimes\rho^*$ and define the symmetric logarithmic derivative by $\partial_\mu\rho=\frac12\{\rho,L_\mu\}$. For each Pauli sector introduce $\mathcal R_P=\sqrt\rho\,P\sqrt\rho$ and $\mathcal T_{\mu,P}=\sqrt\rho\,\{L_\mu,P\}\sqrt\rho$. The Bell statistics satisfy $q^{\rm conj}(P)=d^{-1}\Tr(\mathcal R_P^2)$ and $\partial_\mu q^{\rm conj}(P)=d^{-1}\Tr(\mathcal R_P\mathcal T_{\mu,P})$. At regular points of the Bell probability model, the following decomposition can be applied sector by sector. If a Bell probability vanishes, the corresponding Fisher contribution is understood through the continuous metric extension from nearby regular points, in direct analogy with the treatment of vanishing Pauli probabilities in Appendix~\ref{app:zeros}.

These relations expose the geometry of the information loss. Each Bell probability is the squared Hilbert--Schmidt radius of $\mathcal R_P$, and its derivative detects only the component of $\mathcal T_{\mu,P}$ along that radial direction. Motion orthogonal to $\mathcal R_P$ changes the quantum state but leaves that Bell probability unchanged to first order.

On the regular support, decompose $\mathcal T_{\mu,P}=c_{\mu,P}\mathcal R_P+\mathcal T_{\mu,P}^{\perp}$, with $\Tr(\mathcal R_P\mathcal T_{\mu,P}^{\perp})=0$. The first component is Bell-visible and the second Bell-invisible. With the continuous prescription specified above at nodal sectors, Pauli completeness then gives the exact information gap
\begin{equation}
\left[
F^{\rm Q}[\Omega]
-
F_{\rm Bell}
\right]_{\mu\nu}
=
\frac1d\sum_P
\Tr\!\left(
\mathcal T_{\mu,P}^{\perp}
\mathcal T_{\nu,P}^{\perp}
\right),
\label{eq:mixedgap}
\end{equation}
Equation~\eqref{eq:mixedgap} does more than bound the fixed readout: it identifies the missing information. The Bell histogram records the longitudinal motion of each resolved operator sector, whereas the doubled-state QFIM also contains the transverse motion.

The right-hand side is a positive-semidefinite Gram matrix, and therefore $F_{\rm Bell}\preceq F^{\rm Q}[\Omega]=2F^{\rm Q}[\rho]$. The complete projection derivation and equality conditions are given in Appendix~\ref{app:mixedproof}.

Rank one closes the gap in every regular sector, and the result extends
through nodal sectors by the continuous Fisher-metric prescription
described above. If $\rho^2=\rho$, then $\mathcal R_P=a_P\rho$ and $\mathcal T_{\mu,P}=2(\partial_\mu a_P)\rho$. The resolved motion is therefore entirely longitudinal, $\mathcal T_{\mu,P}^{\perp}=0$, which is the operator-space reason the pure-state Bell identity is exact.

For a full-rank mixed state the opposite conclusion holds. Saturation along a parameter direction $v$ requires every transverse component to vanish; the identity Pauli sector then forces the corresponding SLD to vanish and hence $\partial_v\rho=0$. A nontrivial full-rank trajectory therefore cannot saturate the doubled-state QFI with this fixed Bell readout. Rank-deficient mixed states can evade this conclusion when all parameter dependence remains confined to Bell-visible directions; Appendix~\ref{app:mixedexamples} gives an explicit mixed-spectator construction.

\subsection{A mixed qubit: the gap without many-body complexity}

The transverse loss already occurs for a single full-rank qubit. Consider a Bloch vector of fixed radius $0<r<1$ rotating in the $X$--$Z$ plane,
\begin{equation}
\rho_{r,\vartheta}
=
\frac12
\left[
I+r(\sin\vartheta\,X+\cos\vartheta\,Z)
\right].
\label{eq:mixedQubitFamily}
\end{equation}
Its one-copy QFI is $F_{\vartheta\vartheta}^{\rm Q}[\rho]=r^2$, so the conjugate pair contains $2r^2$. The Bell probabilities are derived in Appendix~\ref{app:mixedqubit}, and their Fisher information is
\begin{equation}
F_{\rm Bell}(r,\vartheta)
=
\frac{
2r^4\sin^2(2\vartheta)
}{
1-r^4+r^4\sin^2(2\vartheta)
}.
\label{eq:mixedQubitBellFI}
\end{equation}
For every $0<r<1$, the Bell Fisher information is strictly below $2r^2$ at generic $\vartheta$. At $\sin(2\vartheta)=0$ it vanishes even though $F_{\vartheta\vartheta}^{\rm Q}[\rho]=r^2$: the state continues to move, but its motion is then entirely transverse to the probability directions resolved by the fixed Bell measurement.

\begin{figure}[!ht]
\centering
\includegraphics[width=\columnwidth]
{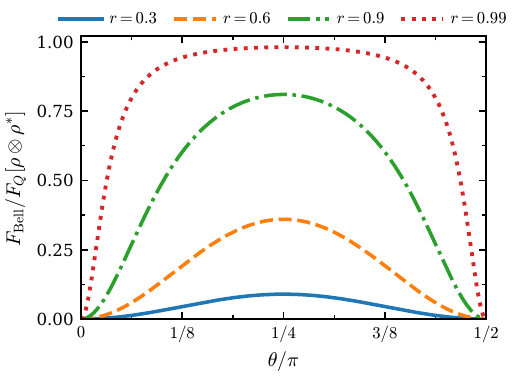}
\caption{\textbf{Mixed-state Bell-information loss.}
Fraction of the doubled-state QFI recovered by the conjugate-copy Bell readout for the fixed-radius family in Eq.~\eqref{eq:mixedQubitFamily}. Finite mixedness produces a parameter-dependent transverse information gap, while the rank-one limit recovers the complete available QFI at regular points.}
\label{fig:mixedQubit}
\end{figure}

As $r\rightarrow1$, the transverse contribution collapses and the Bell-visible fraction approaches unity at regular points. Figure~\ref{fig:mixedQubit} displays this recovery together with the parameter-dependent deficit at finite mixedness.

\subsection{The mixed-state boundary beyond Pauli}

The longitudinal--transverse decomposition is not specific to Pauli strings. For any Hermitian Hilbert--Schmidt orthonormal basis $\mathcal A=\{A_a\}$ introduced in Sec.~\ref{sec:frames}, the physical mixed-state conjugate-pair probabilities are $q_{\mathcal A}(a)=\Tr(A_a\rho A_a\rho)$ rather than the squared coordinates $\Xi_{\mathcal A}(a)=[\Tr(A_a\rho)]^2$. The same operator-space projection gives $F[q_{\mathcal A}]\preceq F^{\rm Q}[\rho\otimes\rho^*]$, with the exact deficit again equal to the Gram matrix of the transverse components. The arbitrary-dimensional expression is derived in Appendix~\ref{app:mixedproof}. The Bell-information gap is therefore a property of mixed-state operator geometry, not of the Pauli group.

The normalized squared-coordinate distribution follows a different geometry. Writing $\widetilde\rho=\rho/\sqrt R$, its Fisher matrix is $F_{\mu\nu}[\Xi_\rho^{\HS}]=4\Tr[(\partial_\mu\widetilde\rho)(\partial_\nu\widetilde\rho)]$. This is the Hilbert--Schmidt geometry of the normalized density operator, generally distinct from both the mixed-state QFIM and the Fisher matrix of the physical conjugate-pair readout. Neither it nor the SLD QFIM universally dominates the other. Its expanded and spectral forms are given in Appendix~\ref{app:mixeddictionary}; the distinction is also relevant to mixed-state extensions of stabilizer R\'enyi constructions \cite{LeoneOlivieroHamma2024Purity,Stratton2026Purity,EspositoViscardiHamma2026Mixed}.

Purity is therefore the boundary at which the squared-coordinate and physical Bell geometries coincide. Inside the mixed-state manifold they separate, and Eq.~\eqref{eq:mixedgap} identifies their physical separation with quantum-state motion transverse to the directions visible in the fixed Bell histogram.

\section{Using Bell data for sensing and quantum geometry}
\label{sec:operational}

The preceding sections determine what information the Bell/Pauli record contains. We now state how that record can be used without introducing an intermediate state-reconstruction step. Two complementary routes are particularly direct. Nearby histograms provide access to the local quantum metric, while a calibrated parametric model turns the same labeled outcomes into a likelihood for sensing and parameter inference.

\subsection{From sampled histograms to quantum geometry}

Suppose a pure-state family can be prepared at two nearby control values,
$\boldsymbol\theta$ and $\boldsymbol\theta+\epsilon v$, and let
$B_{\rm P}(\epsilon)$ be the Bhattacharyya coefficient of the corresponding Bell/Pauli distributions. Equation~\eqref{eq:HellingerLocal} implies
\(F_v^{\rm Q}[\rho]
=
4\lim_{\epsilon\to0}[1-B_{\rm P}(\epsilon)]/\epsilon^2\).
A directional component of the quantum metric can therefore be obtained directly from the classical separation of nearby labeled histograms.

The same construction applies whether the distributions are obtained experimentally or from a numerical Pauli sampler. Repeating it along independent tangent directions, and combining directional curvatures by polarization, reconstructs the local metric tensor on a chosen parameter manifold. The procedure requires neither the underlying state vector nor an intermediate quantum-state reconstruction. In an implementation, finite-$\epsilon$ corrections, sampling fluctuations, and sectors crossing zero must be controlled; the regular limiting prescription for vanishing probabilities is discussed in Appendix~\ref{app:zeros}.

The post-processing itself is not restricted to qubits. For the Hermitian-basis readouts of Sec.~\ref{sec:frames}, the measured outcome is an operator label $a$ rather than a Pauli string $P$, and the same histogram-based construction applies to the resulting probability model. Thus any implementation of the real-coordinate readout provides direct access to the corresponding pure-state metric through classical distribution geometry.

\subsection{From labeled outcomes to parameter inference}

For sensing, one need not first estimate the complete $4^N$-component probability vector. Each Bell shot returns a Pauli-string label $P$. If the experiment is described by a calibrated model
$q^{\rm conj}(P;\boldsymbol\theta)$, observed counts $\{n_P\}$ define the multinomial likelihood
\(\mathcal L_{\rm Bell}(\boldsymbol\theta|\{n_P\})
\propto
\prod_P[q^{\rm conj}(P;\boldsymbol\theta)]^{n_P}\).
For a pure conjugate pair,
$q^{\rm conj}(P;\boldsymbol\theta)
=\Xi_{\rho_{\boldsymbol\theta}}(P)$,
so parameter estimation can be performed directly from the measured labels without quantum-state tomography. For a low-dimensional parametric model, the likelihood only requires evaluation of the probabilities associated with the observed outcomes rather than explicit reconstruction of the full probability vector.

The same record supports score-based estimation of local sensitivity. Whenever
$\partial_\mu\log q^{\rm conj}(P;\boldsymbol\theta)$
can be evaluated, averages over the sampled labels provide estimates of the score covariance and hence of the classical Fisher matrix. This allows local sensitivity to be inferred directly from samples even when storing or reconstructing the full histogram is impractical.

For $\nu$ independent Bell shots, the usual regularity and local-identifiability assumptions give
\(\operatorname{Cov}(\hat{\boldsymbol\theta})
\succeq
\nu^{-1}F_{\rm Bell}^{-1}\).
In the pure conjugate-pair regime,
$F_{\rm Bell}=2F^{\rm Q}[\rho]$.
Since each Bell shot consumes two parameter-dependent registers, a total input budget $N_{\rm in}=2\nu$ gives the one-copy QFIM as the asymptotic Fisher rate per input register. The significance of the construction is therefore not a factor-of-two gain per copy, but access to the complete matrix sensitivity through a fixed readout.

For mixed states the same likelihood machinery applies to the physically sampled $q^{\rm conj}$, with the attainable information given by the Bell-visible Fisher matrix of Sec.~\ref{sec:mixedgap}. Likewise, a signed reconstruction of $q^{\rm conj}$ from identical-copy statistics, as discussed in Sec.~\ref{sec:mixeddistributions}, must retain the transformed sampling law and cannot be treated as direct conjugate-copy data.

Appendix~\ref{app:operationalcomparison} gives an explicit equal-input-copy comparison between a simple local readout, a locally tailored SLD measurement, and the fixed Bell readout. It illustrates how the labeled outcomes enter an ordinary likelihood and how the corresponding Fisher information determines its local width.

\section{Discussion}
\label{sec:discussion}

The main result establishes the complete labeled Pauli distribution as a local statistical representation of pure-state quantum geometry. Although squaring the Pauli coordinates removes discrete sign information and therefore prevents generic finite-distance state reconstruction, it preserves the differential metric exactly: the Fisher--Rao geometry of the distribution is fixed by the pure-state QFIM. This places scalar stabilizer R\'enyi quantities within a broader hierarchy. Their response probes selected contractions of a label-resolved probability flow whose complete Fisher geometry contains strictly more information. In particular, entropy stationarity need not imply either distributional or quantum-state stationarity.

The conjugate-pair construction identifies the structural origin of this correspondence. Vectorization maps a pure density operator to $|\psi\rangle|\psi^*\rangle$, while Hermiticity provides a real coordinate representation in the doubled Hilbert space. The Pauli basis is especially useful because this real-coordinate structure is accompanied by unitarity and tensor-product factorization, turning the corresponding global measurement into pairwise Bell readouts. The higher-dimensional construction shows, however, that neither the Pauli group nor maximal entanglement of every measurement vector is fundamental to Fisher saturation. This provides a constructive operator-space realization of the reality conditions associated with target-independent multiparameter measurements and antiunitary symmetry \cite{MiyazakiMatsumoto2022,Wang2024AntiunitaryQCRB}.

The identical-copy and mixed-state analyses expose two distinct mechanisms by which the pure conjugate-pair correspondence can fail. For identical pure copies, part of the quantum response can remain in phases of the measurement amplitudes, and the available multiparameter Fisher information is constrained even when additional replicas are introduced. The signed relation between identical- and conjugate-copy Bell probabilities sharpens this distinction: the ideal conjugate distribution can be reconstructed algebraically from exact identical-copy probabilities, but this does not reproduce the statistical experiment that generated it. At finite sample number, the transformation acts on the fluctuations together with the mean, so reconstruction of the target probability vector cannot restore Fisher information absent from the original measurement. Algebraic equivalence of probability vectors and statistical equivalence of measurements are therefore distinct notions.

Mixed states exhibit a different obstruction. The normalized squared operator coordinates cease to coincide with the physical conjugate-pair probabilities, and even direct conjugate-copy sampling can fail to resolve the full doubled-state QFIM. The decomposition in Eq.~\eqref{eq:mixedgap} identifies the missing information geometrically as the transverse component of the SLD-resolved operator motion relative to the directions probed by the Bell probabilities. Rank one eliminates this transverse motion sector by sector, whereas nontrivial full-rank trajectories necessarily retain a finite gap. Structured rank-deficient examples show that purity of the complete density operator is not itself the fundamental criterion; the relevant condition is whether the parameter-sensitive motion remains confined to the Bell-visible operator directions.

These results also delimit the metrological meaning of the construction. The factor of two carried by the pure conjugate-pair Fisher matrix reflects two parameter-dependent inputs and does not represent an enhancement of Fisher information per copy. The distinctive feature is instead the measurement architecture: when the conjugated family is physically available, a single fixed readout converts the complete local metric of the doubled pure-state model into an ordinary labeled classical distribution. The same record can then be processed using standard tools of statistical inference without adapting the quantum measurement to an individual parameter direction or reconstructing the state itself.

Several questions remain open at the level relevant for many-body and sensing applications. The present identities specify the information contained in the ideal labeled record, but not the finite-sample cost of extracting a prescribed metric component from an exponentially large outcome space. Determining this sample complexity under physically motivated
structure---including locality, sparsity, symmetries, and restricted parameter manifolds---is therefore essential. Recent general results relating quantum-learning sample complexity to Fisher information provide a natural point of contact for this problem \cite{KwonLieJiang2026}. A second issue is robustness: imperfect preparation of the conjugated branch and measurement noise can reintroduce complex amplitudes or distort the Bell-visible directions, and their effect on the recovered metric remains to be quantified. Finally, the mixed-state projection picture suggests searching for noisy or rank-deficient manifolds on which the transverse contribution is constrained or vanishes, potentially extending fixed-readout quantum-geometry protocols beyond the pure-state boundary.

\begin{acknowledgments}
M.A.R. acknowledges partial support from CNPq and FAPERJ (Grant No.~E26/210.062/2023). E. A. Ramirez Trino acknowledges support from CNPq (Process No.~141672/2023-4).
\end{acknowledgments}

\clearpage

\appendix

\section{Notation}
\label{app:notation}

We summarize the notation used throughout the manuscript. An explicit argument, such as $(P)$ or $(a)$, denotes a single outcome probability, whereas the same symbol without an argument denotes the complete labeled distribution. We use $\Xi$ for squared operator-coordinate distributions and $q$ for physical Born distributions obtained from doubled-state measurements.

\begin{table}[!ht]
\centering
\small
\renewcommand{\arraystretch}{1.15}
\begin{tabular}{@{}l|l@{}}
\toprule
Symbol & Meaning \\
\hline
$\rho$ & density operator \\
$|\psi\rangle$ & pure-state vector, $\rho=|\psi\rangle\langle\psi|$ \\
$d$ & Hilbert-space dimension \\
$P$ & Pauli string, $P\in\mathcal P_N$ \\
$a_P$ & Pauli coordinate, $a_P=\Tr(P\rho)$ \\
$R$ & purity, $R=\Tr(\rho^2)$ \\
$\Xi_\rho(P)$ & pure-state Pauli probability \\
$\Xi_\rho$ & complete pure-state Pauli distribution \\
$\Xi_\rho^{\HS}(P)$ & normalized HS Pauli probability \\
$\Xi_\rho^{\HS}$ & complete normalized HS distribution \\
$q^{\rm conj}(P)$ & conjugate-copy Bell probability \\
$q^{\rm conj}$ & complete conjugate-copy Bell distribution \\
$q^{\rm id}(P)$ & identical-copy Bell probability \\
$q^{\rm id}$ & complete identical-copy Bell distribution \\
$\Pi_P$ & Pauli-labeled Bell projector \\
$\Omega$ & conjugate-pair state, $\rho\otimes\rho^*$ \\
$\mathbf F[p]$ & Fisher matrix of distribution $p$ \\
$F_{\mu\nu}[p]$ & components of $\mathbf F[p]$ \\
$\mathbf F^{\rm Q}[\rho]$ & quantum Fisher information matrix \\
$F_{\mu\nu}^{\rm Q}[\rho]$ & components of $\mathbf F^{\rm Q}[\rho]$ \\
\bottomrule
\end{tabular}
\caption{\textbf{Notation used in the main text.}
An explicit outcome argument denotes a single probability; omission of the argument denotes the complete distribution. The superscript $\HS$ refers to Hilbert--Schmidt normalization and not to Bell statistics.}
\label{tab:notation}
\end{table}

For the operator-frame generalizations, we use
$\Xi_{\mathcal A}(a)$ and $\Xi_{\mathcal A}$ for the single-outcome probability and complete distribution associated with a Hermitian Hilbert--Schmidt orthonormal basis $\mathcal A$, respectively. Likewise,
$q_{\mathcal A}(a)$ and $q_{\mathcal A}$ denote the corresponding physical conjugate-copy probabilities, while
$\Gamma_{\mathcal B}(a)$ and $\Gamma_{\mathcal B}$ denote the squared-coordinate probabilities associated with a generic complex operator basis $\mathcal B$.

\section{Pure-state geometry and entropy response}
\label{app:puregeometry}
This appendix collects the technical statements supporting Sec.~\ref{sec:pure-fisher}. Appendix~\ref{app:zeros} establishes the continuous Fisher metric at vanishing Pauli probabilities; Appendix~\ref{app:finitegeometry} gives the differential-geometric and finite-distance forms of the pure-state identity; Appendix~\ref{app:renyi-response} derives the R\'enyi-response bound of Sec.~\ref{sec:renyi-compression}; and Appendix~\ref{app:shannon-bound} treats its Shannon limit.

\subsection{Regularity at vanishing Pauli probabilities}
\label{app:zeros}

The derivation of Eq.~\eqref{eq:PauliFisherTheorem} is immediate on sectors with $\Xi_\rho(P)>0$, but the standard score representation of the Fisher matrix is singular when a Pauli coordinate vanishes. This subsection shows that the metric itself remains finite and that the pure-state identity extends continuously through such points. With $\Xi_\rho(P)=a_P^2/d$, the regular-sector contribution is
\begin{equation*}
\frac{(\partial_\mu\Xi_\rho(P))(\partial_\nu\Xi_\rho(P))}{\Xi_\rho(P)}=\frac4d(\partial_\mu a_P)(\partial_\nu a_P),
\end{equation*}
which is immediate when $a_P\ne0$. At a point where $a_P=0$, the score $\partial_\mu\log\Xi_\rho(P)$ is not defined. The Fisher metric is instead obtained from the continuous quadratic term of the corresponding statistical distance.

Consider a smooth one-dimensional path $\rho_\epsilon=|\psi_\epsilon\rangle\langle\psi_\epsilon|$ through a point at which $a_P(0)=0$. If $a_P(\epsilon)=\alpha_P\epsilon+O(\epsilon^2)$, then $\Xi_{\rho_\epsilon}(P)=\alpha_P^2\epsilon^2/d+O(\epsilon^3)$.

The one-sided limiting Fisher contribution at nonzero $\epsilon$ is $4\alpha_P^2/d+O(\epsilon)$, so the continuous metric extension is finite. In several parameters, one may define the quadratic form direction by direction and polarize it; the result is again Eq.~\eqref{eq:PauliFisherTheorem}. Nonregular models in which the support changes non-smoothly require the usual care familiar from classical asymptotic theory.

Thus zeros of individual Pauli probabilities do not invalidate the metric identity for a smooth pure-state family; they only require replacing the pointwise score formula by its continuous metric extension.

\subsection{Local and finite-distance geometry of the Pauli-square map}
\label{app:finitegeometry}

This subsection collects the geometric details behind Sec.~\ref{sec:local-global}. The main text uses only the physical statement that nearby pure states and their labeled Pauli distributions carry the same local metric, up to a fixed scale. We first give the compact differential-geometric form of that statement and then derive the finite-distance overlap relation explicitly.

\subsubsection{Formal metric statement}

The manifold of pure states may be represented by the projective space $\mathbb{CP}^{d-1}$, while a normalized probability distribution belongs to the probability simplex $\Delta_{d^2-1}$. The squared-Pauli construction defines the map
\[
\mathcal X:\mathbb{CP}^{d-1}\longrightarrow\Delta_{d^2-1},\qquad[\psi]\mapsto\Xi_\rho.
\]
If $g_{\rm FR}$ denotes the Fisher--Rao metric of the probability distribution and $g$ the pure-state quantum metric, then Eq.~\eqref{eq:PauliFisherTheorem}, together with $F^{\rm Q}_{\mu\nu}=4g_{\mu\nu}$, gives
\begin{equation}
\mathcal X^*g_{\rm FR}=8g.
\label{eq:pullback}
\end{equation}
Thus the Pauli-square map preserves the infinitesimal pure-state geometry up to the constant factor $8$. The notation $\mathcal X^*g_{\rm FR}$ is the pullback of the probability-space metric to the pure-state manifold; no differential-geometric machinery beyond Eq.~\eqref{eq:PauliFisherTheorem} is needed for the results used in the main text.

Equation~\eqref{eq:pullback} also fixes the local identifiability relation between the two descriptions. If the QFIM is nonsingular on a chosen parameter submanifold, then the continuously extended Pauli Fisher matrix is nonsingular on the same submanifold. This local statement does not remove finite-distance ambiguities associated with the signs of the Pauli coordinates or with periodic parameterizations.

\subsubsection{Finite-distance overlap and sign structure}

The Fisher identity is local, but the same Pauli-coordinate representation gives a useful finite-distance statement. Let $\rho=|\psi\rangle\langle\psi|$ and $\sigma=|\varphi\rangle\langle\varphi|$ be two pure states and write $r_P(\rho)=\Tr(P\rho)/\sqrt d$ and $r_P(\sigma)=\Tr(P\sigma)/\sqrt d$. Their Pauli probability distributions are $\Xi_\rho(P)=r_P(\rho)^2$ and $\Xi_\sigma(P)=r_P(\sigma)^2$. The Pauli-distribution overlap is therefore
\begin{equation}
B_{\rm P}(\rho,\sigma)=\sum_P\sqrt{\Xi_\rho(P)\Xi_\sigma(P)}=\sum_P|r_P(\rho)r_P(\sigma)|.
\label{eq:BPappendix}
\end{equation}
Pauli completeness also gives
\begin{equation}
\sum_P r_P(\rho)r_P(\sigma)=\Tr(\rho\sigma)=|\langle\psi|\varphi\rangle|^2.
\label{eq:signedOverlapAppendix}
\end{equation}
Since the last quantity is nonnegative, the triangle inequality yields
\begin{equation}
B_{\rm P}(\rho,\sigma)\geq\Tr(\rho\sigma).
\label{eq:BPglobalBound}
\end{equation}

The distinction between the two sides is entirely due to signs. The state overlap contains the signed products $r_P(\rho)r_P(\sigma)$, whereas $B_{\rm P}$ contains their absolute values. If all nonzero Pauli coordinates retain their signs between the two states, then
\begin{equation}
B_{\rm P}(\rho,\sigma)=\Tr(\rho\sigma).
\label{eq:signChamberEquality}
\end{equation}
A difference between the two quantities can therefore appear only after at least one Pauli coordinate passes through zero.

The local relation used in the main text can also be obtained directly from the probability distributions, without invoking the standard fidelity expansion. Consider any regular normalized distribution $p_\ell$ and a nearby one $p'_\ell=p_\ell+\delta p_\ell$. For each outcome,
\[
\sqrt{p_\ell p'_\ell}=p_\ell\sqrt{1+\frac{\delta p_\ell}{p_\ell}}=p_\ell+\frac12\delta p_\ell-\frac18\frac{(\delta p_\ell)^2}{p_\ell}+O(\delta p^3).
\]
Because both distributions are normalized, $\sum_\ell\delta p_\ell=0$, and therefore
\[
\sum_\ell\sqrt{p_\ell p'_\ell}=1-\frac18\sum_\ell\frac{(\delta p_\ell)^2}{p_\ell}+O(\delta p^3).
\]
For a smooth parameter displacement, $\delta p_\ell=(\partial_\mu p_\ell)\dd\theta^\mu+O(\dd\theta^2)$, so
\begin{equation}
1-B(p_{\boldsymbol\theta},p_{\boldsymbol\theta+\dd\boldsymbol\theta})=\frac18F_{\mu\nu}[p]\,\dd\theta^\mu\dd\theta^\nu+O(\|\dd\boldsymbol\theta\|^3).
\label{eq:classicalBhattacharyyaExpansion}
\end{equation}
Applying Eq.~\eqref{eq:classicalBhattacharyyaExpansion} to the Pauli distribution and using $F_{\mu\nu}[\Xi_\rho]=2F^{\rm Q}_{\mu\nu}[\rho]$ immediately gives
\begin{equation}
B_{\rm P}=1-\frac14F^{\rm Q}_{\mu\nu}[\rho_{\boldsymbol\theta}]\dd\theta^\mu\dd\theta^\nu+O(\|\dd\boldsymbol\theta\|^3).
\label{eq:BhattacharyyaQFI}
\end{equation}
This explicitly explains the factor $1/4$ in Eq.~\eqref{eq:HellingerLocal}: the ordinary overlap of two nearby probability distributions contains one eighth of their Fisher metric, while the Pauli Fisher matrix itself equals twice the pure-state QFIM.

For completeness, the same local metric also appears in the R\'enyi divergence between two nearby distributions, $D_\alpha(p\|q)=(\alpha-1)^{-1}\log\sum_\ell p_\ell^\alpha q_\ell^{1-\alpha}$. For $q=p(\boldsymbol\theta+\dd\boldsymbol\theta)$,
\[
D_\alpha(p_{\boldsymbol\theta}\|p_{\boldsymbol\theta+\dd\boldsymbol\theta})=\frac\alpha2F_{\mu\nu}[p]\,\dd\theta^\mu\dd\theta^\nu+O(\dd\theta^3),
\]
and hence
\begin{equation}
D_\alpha(\Xi_{\rho_{\boldsymbol\theta}}\|\Xi_{\rho_{\boldsymbol\theta+\dd\boldsymbol\theta}})=\alpha F^{\rm Q}_{\mu\nu}[\rho_{\boldsymbol\theta}]\dd\theta^\mu\dd\theta^\nu+O(\dd\theta^3).
\label{eq:RenyiDivergenceQFI}
\end{equation}
This R\'enyi divergence compares two nearby Pauli distributions and should not be confused with the stabilizer R\'enyi entropy, which is a scalar moment of a single distribution.

\subsection{R\'enyi response along a parameter path}
\label{app:renyi-response}

Section~\ref{sec:renyi-compression} uses the response of a single R\'enyi entropy to constrain the QFI along the same deformation of the state. Here we derive Eq.~\eqref{eq:SREgradientBound} directly from the Fisher information of the complete labeled distribution. The calculation isolates the loss produced by replacing the full set of Pauli-sector responses with one R\'enyi moment.

Consider a path $\boldsymbol\theta=\boldsymbol\theta(\lambda)$ with tangent $v^\mu=\dd\theta^\mu/\dd\lambda$ and directional derivative $\partial_v=v^\mu\partial_\mu$. For compactness define
\begin{align*}
Z_\alpha&=\sum_P\Xi_\rho(P)^\alpha,\\
\pi_\alpha(P)&=\frac{\Xi_\rho(P)^\alpha}{Z_\alpha},\\
s_v(P)&=\partial_v\log\Xi_\rho(P)
\end{align*}
on the positive-probability support. Here $\pi_\alpha(P)$ is the normalized weight entering the derivative of the R\'enyi moment, while $s_v(P)$ is the relative rate at which the probability of sector $P$ changes along the path. Since $H_\alpha=(1-\alpha)^{-1}\log Z_\alpha$, one obtains
\begin{equation}
\partial_vH_\alpha=\frac{\alpha}{1-\alpha}\sum_P\pi_\alpha(P)s_v(P).
\label{eq:RenyiDirectionalResponseAppendix}
\end{equation}

Equation~\eqref{eq:RenyiDirectionalResponseAppendix} is the precise compression step: the complete distribution contains the sector-resolved responses $s_v(P)$, whereas $\partial_vH_\alpha$ retains only one weighted combination of them.

Writing
\[
\sum_P\pi_\alpha(P)s_v(P)=\sum_P\left[\sqrt{\Xi_\rho(P)}\,s_v(P)\right]\left[\frac{\Xi_\rho(P)^{\alpha-\frac12}}{Z_\alpha}\right],
\]
Cauchy--Schwarz gives
\[
\left|\sum_P\pi_\alpha(P)s_v(P)\right|^2\leq\left[\sum_P\Xi_\rho(P)s_v(P)^2\right]\frac{Z_{2\alpha-1}}{Z_\alpha^2}.
\]
The first factor is the Fisher information of the complete Pauli distribution along the same path,
\[
\sum_P\Xi_\rho(P)s_v(P)^2=F_v[\Xi_\rho]=2F_v^{\rm Q}[\rho],
\]
where the last equality follows from Eq.~\eqref{eq:PauliFisherTheorem}. Therefore
\begin{equation}
|\partial_vH_\alpha|^2\leq\frac{2\alpha^2}{(1-\alpha)^2}\frac{Z_{2\alpha-1}}{Z_\alpha^2}F_v^{\rm Q}[\rho].
\label{eq:RenyiResponseUpperAppendix}
\end{equation}
Using $Z_{2\alpha-1}/Z_\alpha^2=e^{2(1-\alpha)(H_{2\alpha-1}-H_\alpha)}$, Eq.~\eqref{eq:RenyiResponseUpperAppendix} can be rearranged as
\begin{equation}
F_v^{\rm Q}[\rho]\geq\frac{(1-\alpha)^2}{2\alpha^2}e^{-2(1-\alpha)\left(H_{2\alpha-1}-H_\alpha\right)}|\partial_vH_\alpha|^2,
\end{equation}
which is Eq.~\eqref{eq:SREgradientBound} of the main text.

For $\alpha>1/2$, the expression is well defined on the positive-probability support. At $\alpha=1/2$, $Z_0$ counts the nonzero Pauli weights. For $0<\alpha<1/2$, vanishing probabilities require an explicit regularization because negative powers of the probabilities appear. The case $\alpha=1$ is treated separately in Appendix~\ref{app:shannon-bound}.

\subsection{Shannon response bound}
\label{app:shannon-bound}
The R\'enyi-response derivation of Appendix~\ref{app:renyi-response} does not directly cover $\alpha=1$, where the parametrization in terms of $(1-\alpha)^{-1}$ becomes singular. This subsection derives the corresponding bound for the Shannon entropy and shows that the same one-sided relation between entropy response and QFI survives in this limit.

For
\[
H_1(\Xi_\rho)=-\sum_P\Xi_\rho(P)\log\Xi_\rho(P),
\]
consider the same path $\boldsymbol\theta(\lambda)$ and define $s_v(P)=\partial_v\log\Xi_\rho(P)$ on the positive-probability support. Normalization gives $\sum_P\Xi_\rho(P)s_v(P)=\sum_P\partial_v\Xi_\rho(P)=0$. Therefore
\[
\partial_v H_1=-\sum_P\Xi_\rho(P)s_v(P)\left[\log\Xi_\rho(P)-\langle\log\Xi_\rho\rangle_{\Xi_\rho}\right].
\]
Applying Cauchy--Schwarz gives
\[
|\partial_vH_1|^2\leq\left(\sum_P\Xi_\rho(P)s_v(P)^2\right)\operatorname{Var}_{\Xi_\rho}\!\left(\log\Xi_\rho(P)\right).
\]
The first factor is the directional Fisher information of the complete Pauli distribution, $\sum_P\Xi_\rho(P)s_v(P)^2=F_v[\Xi_\rho]=2F_v^{\rm Q}[\rho]$, where Eq.~\eqref{eq:PauliFisherTheorem} was used in the last step. Hence
\begin{equation}
|\partial_vH_1|^2\leq2F_v^{\rm Q}[\rho]\,\operatorname{Var}_{\Xi_\rho}\!\left(\log\Xi_\rho(P)\right).
\label{eq:ShannonGradientBound}
\end{equation}

Equation~\eqref{eq:ShannonGradientBound} is the Shannon counterpart of Eq.~\eqref{eq:SREgradientBound}. As in the R\'enyi case, the implication is one-sided: a nonzero Shannon response requires motion of the labeled Pauli distribution and constrains its QFI, whereas compensating motion among Pauli sectors may leave $H_1$ stationary.

\section{Bell-basis construction and explicit qubit checks}
\label{app:qubitchecks}

{\color{black}
This appendix gives the algebraic details behind the Bell construction of Sec.~\ref{sec:bell}. We first show how the Pauli operator basis becomes a Bell basis in the doubled Hilbert space and why the global $N$-qubit measurement factorizes into independent measurements between corresponding sites. We then give explicit one-qubit checks of the probabilities and factors of two.
}

\subsection{From Pauli operators to Bell states}

{\color{black}
For an operator
}
\[
A
=
\sum_{ij}A_{ij}|i\rangle\langle j|,
\]
{\color{black}
we use column vectorization,
}
\[
\vecop{A}
=
\sum_{ij}
A_{ij}|i\rangle_A|j\rangle_B.
\]
{\color{black}
The ordinary operator inner product becomes
}
\[
\langle\!\langle A|B\rangle\!\rangle
=
\Tr(A^\dagger B).
\]
{\color{black}
For a pure projector
$\rho=|\psi\rangle\langle\psi|$,
}
\[
\vecop{\rho}
=
|\psi\rangle_A|\psi^*\rangle_B.
\]

{\color{black}
Pauli orthogonality,
$\Tr(PP')=d\,\delta_{P,P'}$,
implies that the normalized vectors
$\vecop{P}/\sqrt d$
form an orthonormal basis of the doubled Hilbert space. The corresponding projectors are
}
\[
\Pi_P
=
\frac{\vecop{P}\brvec{P}}{d},
\qquad
\sum_P\Pi_P=I.
\]
{\color{black}
For a general two-register density operator $\Omega$, Born's rule gives
}
\[
\Pr(P|\Omega)
=
\Tr(\Pi_P\Omega).
\]
{\color{black}
For the pure conjugate pair,
}
\[
\Omega
=
\rho\otimes\rho^*
=
\vecop{\rho}\brvec{\rho},
\]
{\color{black}
so
}
\[
\begin{aligned}
\Pr(P|\rho\otimes\rho^*)
&=
\frac1d
|\langle\!\langle P|\rho\rangle\!\rangle|^2
\\
&=
\frac1d
|\Tr(P\rho)|^2.
\end{aligned}
\]
{\color{black}
Because $P$ and $\rho$ are Hermitian,
$\Tr(P\rho)$ is real, and hence
}
\[
\Pr(P|\rho\otimes\rho^*)
=
\frac1d[\Tr(P\rho)]^2
=
\Xi_\rho(P),
\]
{\color{black}
which reproduces Eq.~\eqref{eq:BellXi}.
}

{\color{black}
For one qubit, the connection with the usual Bell basis is explicit:
}
\[
\begin{array}{c@{\qquad}c}
\dfrac{\vecop{I}}{\sqrt2}=|\Phi^+\rangle,
&
\dfrac{\vecop{Z}}{\sqrt2}=|\Phi^-\rangle,
\\[2mm]
\dfrac{\vecop{X}}{\sqrt2}=|\Psi^+\rangle,
&
\dfrac{\vecop{Y}}{\sqrt2}=-\ii|\Psi^-\rangle.
\end{array}
\]
{\color{black}
The phase multiplying the $Y$ state has no effect on its measurement projector. Thus the four Bell outcomes can be labeled by
$I,X,Y,Z$.
}

\subsection{Factorization for an $N$-qubit state}

{\color{black}
For a Pauli string
$P=P_1\otimes\cdots\otimes P_N$,
vectorization respects the tensor-product structure after regrouping the two copies site by site:
}
\[
\frac{
\vecop{P_1\otimes\cdots\otimes P_N}
}{
\sqrt{2^N}
}
=
\bigotimes_{j=1}^{N}
\frac{\vecop{P_j}}{\sqrt2}.
\]
{\color{black}
The projector onto the global vectorized Pauli state therefore factorizes as
}
\[
\Pi_P
=
\bigotimes_{j=1}^{N}\Pi_{P_j}.
\]
{\color{black}
Consequently, measuring the global vectorized Pauli basis is physically equivalent to performing an ordinary Bell measurement between each pair of corresponding qubits. This factorization concerns only the readout between the two systems and does not require the many-body state within either system to factorize.
}

\subsection{Explicit one-qubit check}

{\color{black}
We now verify the Bell probabilities and Fisher matrices explicitly for the general one-qubit pure state.
}

Write
\begin{equation*}
 |\psi(\theta,\phi)\rangle
 =
 c_\theta|0\rangle+\e^{\ii\phi}s_\theta|1\rangle,
 \quad
 c_\theta=\cos\frac\theta2,
 \quad
 s_\theta=\sin\frac\theta2.
\end{equation*}
{Let $\rho=|\psi\rangle\langle\psi|$.}
Its Pauli expectations are
\begin{align*}
 \langle I\rangle&=1,&
 \langle Z\rangle&=\cos\theta,\\
 \langle X\rangle&=\sin\theta\cos\phi,&
 \langle Y\rangle&=\sin\theta\sin\phi.
\end{align*}
The Pauli probability spectrum is therefore
\begin{equation}
 {\Xi_\rho}=\frac12\left(
 1,
 \sin^2\theta\cos^2\phi,
 \sin^2\theta\sin^2\phi,
 \cos^2\theta
 \right).
 \label{eq:qubitXiCheck}
\end{equation}
It is normalized because the Bloch vector has unit length.
Away from its zero sectors, direct differentiation gives
\begin{equation*}
 {\mathbf F[\Xi_\rho]}=
 \begin{pmatrix}
 2&0\\
 0&2\sin^2\theta
 \end{pmatrix}
 =
 {2\mathbf F^{\rm Q}[\rho]},
\end{equation*}
with
\begin{equation*}
 {\mathbf F^{\rm Q}[\rho]}=
 \begin{pmatrix}
 1&0\\
 0&\sin^2\theta
 \end{pmatrix}.
\end{equation*}
The identity remains valid at $\phi=0$, where the $Y$ probability vanishes,
when the Fisher metric is interpreted by the regularized limit.

The conjugate-pair state is
\begin{equation*}
 |\psi\rangle|\psi^*\rangle
 =
 c_\theta^2|00\rangle
 +c_\theta s_\theta \e^{-\ii\phi}|01\rangle
 +c_\theta s_\theta \e^{\ii\phi}|10\rangle
 +s_\theta^2|11\rangle.
\end{equation*}
Projection onto the Bell basis gives precisely
Eq.~\eqref{eq:qubitXiCheck}, up to the conventional assignment of
$X,Y,Z$ to the three nontrivial Bell states.
The doubled-state QFIM is
\begin{equation*}
 {\mathbf F^{\rm Q}[\rho\otimes\rho^*]}
 =
 2\begin{pmatrix}
 1&0\\
 0&\sin^2\theta
 \end{pmatrix},
\end{equation*}
so the Bell measurement saturates both parameters simultaneously.

For comparison, two identical copies give
\begin{equation*}
 |\psi\rangle^{\otimes2}
 =
 c_\theta^2|00\rangle
 +c_\theta s_\theta \e^{\ii\phi}(|01\rangle+|10\rangle)
 +s_\theta^2\e^{2\ii\phi}|11\rangle.
\end{equation*}
The antisymmetric Bell outcome has zero probability, and the other Bell
amplitudes contain a relative phase $2\phi$.
Their probabilities are the $n=2$ specialization of
Appendix~\ref{app:weylFI}.
The classical Bell Fisher matrix obeys the normalized trace budget
\begin{equation*}
 \frac{F_{\theta\theta}}{2}+\frac{F_{\phi\phi}}{2\sin^2\theta}=1,
\end{equation*}
rather than equaling the doubled QFIM in both directions.
This elementary example already displays the entire
conjugate-versus-identical distinction.

As a mixed-state check take a qubit
\begin{equation*}
 \rho=\frac12(I+\mathbf r\cdot\boldsymbol\sigma),\qquad |\mathbf r|<1.
\end{equation*}
For $P=I$,
${q^{\rm conj}(I)}=\frac12\Tr\rho^2=(1+|\mathbf r|^2)/4$.
Conjugation by $X,Y,Z$ flips two components of the Bloch vector, so the
remaining Bell probabilities are
\begin{align*}
 {q^{\rm conj}(X)}&=\frac14(1+r_x^2-r_y^2-r_z^2),\\
 {q^{\rm conj}(Y)}&=\frac14(1-r_x^2+r_y^2-r_z^2),\\
 {q^{\rm conj}(Z)}&=\frac14(1-r_x^2-r_y^2+r_z^2).
\end{align*}
These four quantities sum to one and are nonnegative for physical $\rho$.
In the pure limit $|\mathbf r|=1$ they reduce to
\begin{equation*}
 {q^{\rm conj}(P)}=\frac12[\Tr(P\rho)]^2,
\end{equation*}
whereas for $|\mathbf r|<1$ they differ from the normalized squared Bloch
coordinates.
This gives a concrete one-qubit illustration of the three-distribution
distinction in Appendix~\ref{app:mixeddictionary}.

\section{Identical-copy measurements and Fisher-information budget}
\label{app:identicalcopies}

\subsection{Derivation of the phase-covariance identity}
\label{app:phase}

Let $|\psi_{\boldsymbol\theta}\rangle=\sum_a c_a|e_a\rangle$, with $c_a=r_a e^{\ii\phi_a}$. Then
\[
\partial_\mu c_a=e^{\ii\phi_a}(\partial_\mu r_a+\ii r_a\partial_\mu\phi_a).
\]
Normalization gives $\sum_ar_a\partial_\mu r_a=0$. The real part of $\langle\partial_\mu\psi|\partial_\nu\psi\rangle$ is
\[
\sum_a\left[\partial_\mu r_a\partial_\nu r_a+r_a^2\partial_\mu\phi_a\partial_\nu\phi_a\right],
\]
while
\[
\langle\psi|\partial_\mu\psi\rangle=\ii\sum_ar_a^2\partial_\mu\phi_a.
\]
Substitution into the pure-state QFIM immediately yields Eq.~\eqref{eq:phaseCov}. This decomposition is invariant under a parameter-dependent global phase because the covariance subtracts the mean phase velocity.

The same decomposition distinguishes conjugate and identical pairs at the operator level. With the vectorization convention of Sec.~\ref{sec:bell},
\[
|\psi\rangle|\psi^*\rangle=\vecop{\rho},\qquad \rho=|\psi\rangle\langle\psi|,
\]
and the amplitude along a normalized vectorized Pauli operator is
\[
\frac{\langle\!\langle P|\rho\rangle\!\rangle}{\sqrt d}=\frac{\Tr(P\rho)}{\sqrt d}.
\]
Because $P$ and $\rho$ are Hermitian, this amplitude is real and cannot carry parameter-dependent phase motion. For two identical copies, instead,
\[
|\psi\rangle|\psi\rangle=\vecop{\,|\psi\rangle\langle\psi^*|\,}.
\]
The vectorized operator now has matrix elements $c_ic_j$ rather than $c_ic_j^*$ and is generally non-Hermitian. Its Pauli-basis amplitudes can therefore acquire parameter-dependent phases that disappear upon taking their squared magnitudes. This is the operator-space form of the amplitude--phase distinction in Eq.~\eqref{eq:phaseCov}.

\subsection{An elementary identical-copy Fisher-budget proof}
\label{app:GMproof}

We give a local derivation of the pure-state Fisher budget used in Sec.~\ref{sec:identical}. The general identical-copy result is broader \cite{GillMassar2000}; the present derivation fixes the normalization and makes the factor-of-two comparison explicit.

Choose a point $|0\rangle$ on the pure-state manifold in dimension $d$ and local real coordinates $x_a,y_a$, $a=1,\ldots,d-1$, such that
\begin{equation}
|\psi(\mathbf x,\mathbf y)\rangle
=|0\rangle+\frac12\sum_{a=1}^{d-1}(x_a+\ii y_a)|a\rangle
+O(\|(\mathbf x,\mathbf y)\|^2).
\label{eq:localPureCoords}
\end{equation}
At the origin the one-copy QFIM is the identity on the $2(d-1)$ real tangent directions. For $n$ copies, let $|\Psi_n\rangle=|\psi(\mathbf x,\mathbf y)\rangle^{\otimes n}$ and define
\[
|T_a\rangle=\sum_{r=1}^{n}|0\rangle^{\otimes(r-1)}|a\rangle|0\rangle^{\otimes(n-r)},\qquad \langle T_a|T_b\rangle=n\delta_{ab}.
\]
Then
\[
\partial_{x_a}|\Psi_n\rangle=\frac12|T_a\rangle,\qquad \partial_{y_a}|\Psi_n\rangle=\frac{\ii}{2}|T_a\rangle.
\]

An arbitrary POVM may be refined into rank-one outcomes, and such refinement cannot decrease the classical Fisher information. It is therefore sufficient to consider $E_\xi=|m_\xi\rangle\langle m_\xi|$. At a regular point let
\[
c_\xi=\langle m_\xi|\Psi_n\rangle,\qquad t_{\xi a}=\langle m_\xi|T_a\rangle.
\]
The outcome probability and derivatives are
\begin{align*}
p_\xi&=|c_\xi|^2,\\
\partial_{x_a}p_\xi&=\Re(c_\xi^*t_{\xi a}),\\
\partial_{y_a}p_\xi&=-\Im(c_\xi^*t_{\xi a}).
\end{align*}
For the rank-one refinement,
\begin{equation}
\frac{(\partial_{x_a}p_\xi)^2+(\partial_{y_a}p_\xi)^2}{p_\xi}
=|t_{\xi a}|^2.
\label{eq:GMoutcomeCS}
\end{equation}
Summing over the refined outcomes and using POVM completeness gives
\[
F_{x_ax_a}+F_{y_ay_a}=\sum_\xi|t_{\xi a}|^2=\langle T_a|T_a\rangle=n.
\]
A coarse-grained measurement can only reduce the Fisher information, so summing over $a$ yields
\[
\Tr \mathbf F^{(n)}\le n(d-1)
\]
in the local coordinates in which the one-copy QFIM is the identity. Covariance under coordinate changes gives Eq.~\eqref{eq:GM}.

For a qubit there is only one complex tangent direction. A complete rank-one projective measurement saturates Eq.~\eqref{eq:GMoutcomeCS} after summation, and hence
\begin{equation}
\Tr\!\left[
(\mathbf F^{\rm Q,(n)})^{-1}\mathbf F^{(n)}
\right]=1.
\label{eq:rankOneQubitTrace}
\end{equation}
This identity underlies Eq.~\eqref{eq:traceidentity} for the relative-Weyl readout. It constrains the total Fisher budget without requiring its distribution between the two tangent directions to be isotropic.

\subsection{Relative-Weyl geometry in detail}
\label{app:weyl}

We derive the even--odd replica structure of the relative-Weyl measurement. Let $V=\mathbb F_2^2$ carry the usual nondegenerate alternating form $[v,w]$. The $n$-copy Pauli phase space is $V^n\simeq\mathbb F_2^n\otimes V$ with form
\begin{equation}
[(a\otimes v),(b\otimes w)]_n=(a\cdot b)[v,w].
\label{eq:tensorSymplectic}
\end{equation}
The collective one-copy phase-space vector $u\in V$ is represented by $\mathbf1\otimes u$, where $\mathbf1=(1,\ldots,1)$. A relative operator commutes with all collective Weyl operators iff its copy-space component belongs to
\[
U_n=\mathbf1^\perp=\left\{a\in\mathbb F_2^n:\sum_ra_r=0\right\}.
\]
Thus the relative phase space is
\[
V_n^{\rm rel}=U_n\otimes V,
\]
of dimension $2(n-1)$.

The restriction of the ordinary dot product to $U_n$ depends on the parity of $n$. If $n$ is odd, $\mathbf1\notin U_n$, so $U_n\cap U_n^\perp=\{0\}$ and the induced form on $V_n^{\rm rel}$ is nondegenerate. Any maximal isotropic subspace then has dimension $n-1$. The choice
\[
\Lambda_n^{\rm odd}(\ell)=U_n\otimes\ell,\qquad \dim\ell=1,
\]
contains $n-1$ commuting Pauli generators. On $n$ physical qubits this fixes only $n-1$ stabilizer eigenvalues and leaves a two-dimensional logical space. This residual collective qubit is unavoidable if one restricts the measurement algebra to the relative commutant.

If $n$ is even, $\mathbf1\in U_n$. The radical of the restricted form is
\[
\operatorname{rad}(V_n^{\rm rel})=\mathcal K_n=\mathbf1\otimes V,\qquad \dim\mathcal K_n=2.
\]
Choose a one-dimensional Lagrangian line $\ell\subset V$. Then
\[
\Lambda_n(\ell)=\mathcal K_n+(U_n\otimes\ell)
\]
is isotropic. Since $\mathcal K_n\cap(U_n\otimes\ell)=\mathbf1\otimes\ell$,
\[
\dim\Lambda_n=2+(n-1)-1=n.
\]
An isotropic subspace containing the full radical can have at most
\[
\dim\mathcal K_n+\frac12\dim(V_n^{\rm rel}/\mathcal K_n)=2+(n-2)=n
\]
dimensions, so $\Lambda_n$ is maximal. It therefore supplies $n$ independent commuting Pauli stabilizers and a rank-one projective measurement on the $n$-qubit Hilbert space.

For $\ell=\operatorname{span}(Z)$, the $Z$-type relative generators can be taken as $Z_aZ_n$, $a=1,\ldots,n-1$. The radical contributes one new independent generator transverse to this family, which can be chosen as $X^{\otimes n}$; $Z^{\otimes n}$ is already generated by the relative $Z$ parities for even $n$. Thus a convenient generating set is
\begin{equation}
X^{\otimes n},\qquad Z_aZ_n,\quad a=1,\ldots,n-1.
\label{eq:catgens}
\end{equation}

The $Z_aZ_n$ eigenvalues determine a bit string $\mathbf b\in\{0,1\}^n$ modulo global complement, while $X^{\otimes n}$ distinguishes the symmetric and antisymmetric combinations. The corresponding joint eigenstates are
\begin{equation}
|[\mathbf b],\pm\rangle
=
\frac{|\mathbf b\rangle\pm|\bar{\mathbf b}\rangle}{\sqrt2},
\qquad
[\mathbf b]=\{\mathbf b,\bar{\mathbf b}\}.
\label{eq:catstates}
\end{equation}
These are the complement-pair states used in the Fisher analysis of Appendix~\ref{app:weylFI}.

The construction is covariant under permutations of the copies. Permutations preserve $U_n$, fix the radical $\mathcal K_n$, and act trivially on the choice of one-copy polarization $\ell$. A different polarization is related by a one-copy Clifford transformation applied identically to all copies. Hence the construction is canonical up to the choice of one-copy polarization.

The relation with the ordinary two-copy Bell basis is exact. At $n=2$, $U_2=\operatorname{span}(11)$ and $\mathcal K_2=\mathbf1\otimes V=V_2^{\rm rel}$, so a maximal commuting basis is generated by $XX$ and $ZZ$. At $n=4$, the generators are $XXXX,Z_1Z_4,Z_2Z_4,Z_3Z_4$; at $n=6$ they are $X^{\otimes6}$ together with five relative $Z$ parities. The state labels are respectively $2$, $8$, and $32$ complement classes, each with a $\pm$ parity, giving $2^n$ projectors as required.

\subsection{Exact Fisher analysis for $n=2,4,6$ and general even $n$}
\label{app:weylFI}

We now provide the calculation behind Eqs.~\eqref{eq:traceidentity} and \eqref{eq:phasebound}. For one complement class with representative weight $k$, define
\begin{align*}
\omega_k&=c_\vartheta^{\,2(n-k)}s_\vartheta^{\,2k},\\
\bar\omega_k&=c_\vartheta^{\,2k}s_\vartheta^{\,2(n-k)},\\
\Sigma_k&=\omega_k+\bar\omega_k,\\
\kappa_n&=2(c_\vartheta s_\vartheta)^n,\\
\Delta_k&=n-2k,
\end{align*}

where $c_\vartheta=\cos(\vartheta/2)$ and $s_\vartheta=\sin(\vartheta/2)$. The two probabilities are
\begin{equation}
p_{k,\pm}=\frac12\left[\Sigma_k\pm\kappa_n\cos(\Delta_k\varphi)\right].
\label{eq:pkpmDetailed}
\end{equation}
They depend on the particular bit string only through its Hamming weight. For $k<n/2$ there are $\binom nk$ complement classes with representative weight $k$; at $k=n/2$ there are $\frac12\binom n{n/2}$ classes.

The phase derivative is
\[
\partial_\varphi p_{k,\pm}
=\mp\frac12\kappa_n\Delta_k\sin(\Delta_k\varphi),
\]
so the phase-Fisher contribution of one complement class is
\begin{equation}
f_{\varphi}^{(k)}
=
\frac{\kappa_n^2\Delta_k^2\sin^2(\Delta_k\varphi)\,\Sigma_k}
{\Sigma_k^2-\kappa_n^2\cos^2(\Delta_k\varphi)}.
\label{eq:fphik}
\end{equation}
The arithmetic-geometric mean inequality gives $\Sigma_k\ge\kappa_n$. Setting $t=\Sigma_k/\kappa_n\ge1$, one finds
\[
\Sigma_k^2-\kappa_n^2\cos^2\alpha-\kappa_n\Sigma_k\sin^2\alpha
=\kappa_n^2(t-1)(t+\cos^2\alpha)\ge0.
\]
Hence
\begin{equation}
f_\varphi^{(k)}\le \kappa_n\Delta_k^2.
\label{eq:pairPhaseBound}
\end{equation}
Summing over complement classes uses
\[
\sum_{\mathbf b\in\{0,1\}^n}(n-2|\mathbf b|)^2=n2^n.
\]
Every complement pair occurs twice in this sum and has the same $\Delta_k^2$, so
\[
\sum_{[\mathbf b]}\Delta_{|\mathbf b|}^2=n2^{n-1}.
\]
Since $\kappa_n=2^{1-n}\sin^n\vartheta$, Eq.~\eqref{eq:pairPhaseBound} yields
\[
F_{\varphi\varphi}^{(n)}\le\kappa_n n2^{n-1}=n\sin^n\vartheta,
\]
which is Eq.~\eqref{eq:phasebound}. Equality holds at the equator for regular phase points.

The normalized trace identity follows directly from Eq.~\eqref{eq:rankOneQubitTrace}. On the $(\vartheta,\varphi)$ manifold,
\[
\mathbf F^{\rm Q,(1)}
=
\begin{pmatrix}
1&0\\
0&\sin^2\vartheta
\end{pmatrix},
\qquad
\mathbf F^{\rm Q,(n)}
=
n\mathbf F^{\rm Q,(1)}.
\]
Applying Eq.~\eqref{eq:rankOneQubitTrace} to the relative-Weyl basis gives
\[
\frac{F_{\vartheta\vartheta}^{(n)}}{n}+\frac{F_{\varphi\varphi}^{(n)}}{n\sin^2\vartheta}=1,
\]
which is Eq.~\eqref{eq:traceidentity}. The off-diagonal entry $F_{\vartheta\varphi}^{(n)}$ need not vanish, but positivity implies
\[
|F_{\vartheta\varphi}^{(n)}|^2
\le
F_{\vartheta\vartheta}^{(n)}F_{\varphi\varphi}^{(n)},
\]
so suppression of the phase diagonal also suppresses the normalized cross term away from the equator.

Together with Eq.~\eqref{eq:traceidentity}, this gives, for every fixed $\vartheta\neq\pi/2$,
\begin{equation}
\begin{split}
&
\left(\mathbf F^{\rm Q,(1)}\right)^{-1/2}
\frac{\mathbf F^{(n)}}{n}
\left(\mathbf F^{\rm Q,(1)}\right)^{-1/2}
\\
&\hspace{3em}
\longrightarrow
\operatorname{diag}(1,0).
\end{split}
\label{eq:asymptoticFisher}
\end{equation}
Thus the normalized phase and cross components vanish away from the equator, while the population-changing component approaches unity.

For $n=2$ there are two complement classes. The $k=0$ class gives
\[
p_{0,\pm}^{(2)}
=
\frac12\left(c_\vartheta^4+s_\vartheta^4
\pm2c_\vartheta^2s_\vartheta^2\cos2\varphi\right),
\]
while for $k=1$,
\[
p_{1,+}^{(2)}=2c_\vartheta^2s_\vartheta^2,
\qquad
p_{1,-}^{(2)}=0.
\]
The zero probability is the antisymmetric singlet and is handled by the limiting prescription of Appendix~\ref{app:zeros}.

For $n=4$, the class multiplicities for $k=0,1,2$ are $1,4,3$. The nontrivial probabilities are
\begin{align*}
p_{0,\pm}^{(4)}
&=\frac12\left(c_\vartheta^8+s_\vartheta^8
\pm2c_\vartheta^4s_\vartheta^4\cos4\varphi\right),\\
p_{1,\pm}^{(4)}
&=\frac12\left(c_\vartheta^6s_\vartheta^2+c_\vartheta^2s_\vartheta^6
\pm2c_\vartheta^4s_\vartheta^4\cos2\varphi\right),\\
p_{2,+}^{(4)}&=2c_\vartheta^4s_\vartheta^4,
\qquad
p_{2,-}^{(4)}=0.
\end{align*}

The $k=1$ expression occurs four times and the $k=2$ expression three times.

For $n=6$, the multiplicities for $k=0,1,2,3$ are $1,6,15,10$, and
\begin{align*}
p_{0,\pm}^{(6)}
&=\frac12\left(c_\vartheta^{12}+s_\vartheta^{12}
\pm2c_\vartheta^6s_\vartheta^6\cos6\varphi\right),\\
p_{1,\pm}^{(6)}
&=\frac12\left(c_\vartheta^{10}s_\vartheta^2+c_\vartheta^2s_\vartheta^{10}
\pm2c_\vartheta^6s_\vartheta^6\cos4\varphi\right),\\
p_{2,\pm}^{(6)}
&=\frac12\left(c_\vartheta^8s_\vartheta^4+c_\vartheta^4s_\vartheta^8
\pm2c_\vartheta^6s_\vartheta^6\cos2\varphi\right),\\
p_{3,+}^{(6)}&=2c_\vartheta^6s_\vartheta^6,
\qquad
p_{3,-}^{(6)}=0.
\end{align*}

These expressions show directly that all $\varphi$ dependence is controlled by the common interference scale $\kappa_n$.

Finally, for $\vartheta=\pi/2+\delta$ with $|\delta|\ll1$,
\[
\sin^n\vartheta=\left(1-\frac{\delta^2}{2}+O(\delta^4)\right)^n=\exp\left[-\frac{n\delta^2}{2}+O(n\delta^4)\right].
\]
The phase-sensitive region therefore has width $|\delta|=O(n^{-1/2})$. Outside this window the relative-Weyl readout becomes asymptotically concentrated on the population-changing direction, whereas within it the factors $\cos[(n-2k)\varphi]$ retain phase sensitivity.

\section{Hermitian readouts in arbitrary dimension}
\label{app:hermitianreadouts}

This appendix collects the constructions underlying the real-amplitude mechanism of Sec.~\ref{sec:frames}. We first give explicit Hermitian operator bases and their local realizations in arbitrary finite dimension, then extend the construction to Hermitian tight frames and quantify the information lost by generic complex operator coordinates. Finally, we show that within orthogonal operator-basis readouts, universal pure-state Fisher saturation requires Hermiticity up to fixed phase factors.

\subsection{Canonical Hermitian readouts and local realizations}
\label{app:higherdim}

We give explicit realizations of the Hermitian readout used in Sec.~\ref{sec:frames}. Let $\dim\mathcal H_{\rm loc}=d_{\rm loc}$ and choose a reference basis $\{|i\rangle\}_{i=0}^{d_{\rm loc}-1}$. Define
\[
D_i=|i\rangle\langle i|,
\]
and, for $i<j$,
\[
X_{ij}
=
\frac{|i\rangle\langle j|+|j\rangle\langle i|}{\sqrt2},
\qquad
Y_{ij}
=
\frac{-\ii|i\rangle\langle j|+\ii|j\rangle\langle i|}{\sqrt2}.
\]
The $d_{\rm loc}^2$ operators $\{D_i,X_{ij},Y_{ij}\}$ form a Hermitian Hilbert--Schmidt orthonormal basis $\mathcal A_{\rm can}$. The density matrix expands as
\[
\rho
=
\sum_i\rho_{ii}D_i
+
\sqrt2\sum_{i<j}\Re(\rho_{ij})X_{ij}
-
\sqrt2\sum_{i<j}\Im(\rho_{ij})Y_{ij},
\]
so all coordinates $\eta_a=\Tr(A_a\rho)$ are manifestly real. Vectorization gives
\begin{align*}
\vecop{D_i}=|ii\rangle,
\quad
\vecop{X_{ij}}
=
\frac{|ij\rangle+|ji\rangle}{\sqrt2},
\\
\vecop{Y_{ij}}
=
\frac{-\ii|ij\rangle+\ii|ji\rangle}{\sqrt2}.
\end{align*}
For a pure state, projective measurement in this basis samples
\[
\Xi_{\mathcal A_{\rm can}}(a)
=
[\Tr(A_a\rho)]^2,
\]
and therefore has Fisher matrix $2F^{\rm Q}_{\mu\nu}[\rho]$.

For $d_{\rm loc}=2$, an orthogonal rotation in the diagonal sector gives
\[
\frac{D_0+D_1}{\sqrt2}
=
\frac{I}{\sqrt2},
\qquad
\frac{D_0-D_1}{\sqrt2}
=
\frac{Z}{\sqrt2},
\]
while $X_{01}=X/\sqrt2$ and $Y_{01}=Y/\sqrt2$. The canonical Hermitian basis is therefore equivalent, up to this real rotation, to the normalized Pauli basis, and its vectorization gives the ordinary Bell basis.

For $d_{\rm loc}>2$, maximal entanglement of the vectorized basis is no longer required. In particular, $\vecop{D_i}=|ii\rangle$ is a product state. A vectorized operator is maximally entangled only when the operator is proportional to a unitary, whereas Eq.~\eqref{eq:HSFisher} follows from the reality of Hermitian operator coordinates. These two properties coincide for the qubit Pauli basis but separate in higher dimension.

For a qutrit, the nine canonical vectorized basis states are
\begin{equation}
\begin{gathered}
|00\rangle,\quad |11\rangle,\quad |22\rangle,\\
\frac{|01\rangle+|10\rangle}{\sqrt2},
\quad
\frac{|02\rangle+|20\rangle}{\sqrt2},
\quad
\frac{|12\rangle+|21\rangle}{\sqrt2},\\
\frac{-\ii|01\rangle+\ii|10\rangle}{\sqrt2},
\quad
\frac{-\ii|02\rangle+\ii|20\rangle}{\sqrt2},
\quad
\frac{-\ii|12\rangle+\ii|21\rangle}{\sqrt2}.
\end{gathered}
\label{eq:qutritHermitianBasis}
\end{equation}
For a pure qutrit density matrix, the corresponding probability coordinates are
\begin{equation}
\begin{gathered}
\rho_{00}^2,\quad \rho_{11}^2,\quad \rho_{22}^2,\\
2[\Re(\rho_{01})]^2,\quad
2[\Re(\rho_{02})]^2,\quad
2[\Re(\rho_{12})]^2,\\
2[\Im(\rho_{01})]^2,\quad
2[\Im(\rho_{02})]^2,\quad
2[\Im(\rho_{12})]^2.
\end{gathered}
\label{eq:qutritHermitianProbabilities}
\end{equation}
They sum to $\Tr(\rho^2)=1$, and their Fisher matrix therefore equals $2F^{\rm Q}[\rho]$. This example makes explicit that no operator-group structure is required: the measurement resolves and squares a complete set of real coordinates of the Hermitian density matrix.

For a spin-$s$ degree of freedom, $d_{\rm loc}=2s+1$, a rotationally adapted Hermitian basis can be constructed from irreducible spherical tensor operators $T_{KQ}$, with $K=0,\ldots,2s$ and $Q=-K,\ldots,K$. Using $T_{KQ}^\dagger=(-1)^QT_{K,-Q}$, define for $Q>0$
\begin{align*}
C_{KQ}=\frac{T_{KQ}+(-1)^QT_{K,-Q}}{\sqrt2},
\\
S_{KQ}=\frac{T_{KQ}-(-1)^QT_{K,-Q}}{\sqrt2\,\ii},
\end{align*}
together with $T_{K0}$. After Hilbert--Schmidt normalization, these operators form a Hermitian orthonormal basis. Their real coordinates $\eta_a=\Tr(A_a\rho)$ define $\Xi_{\mathcal A}(a)=\eta_a^2$, whose Fisher matrix equals $2F^{\rm Q}[\rho]$ for every pure spin family.

For spin one, the nine Hermitian coordinates may equivalently be chosen from the identity, the three spin components $S_x,S_y,S_z$, and five independent traceless quadrupoles obtained from Hermitian combinations of $S_iS_j$. After Hilbert--Schmidt orthonormalization they form a basis $\mathcal A=\{A_a\}_{a=0}^{8}$ of $\operatorname{Herm}(\mathbb C^3)$, so that
\[
\rho=\sum_{a=0}^{8}\eta_aA_a,
\qquad
\sum_{a=0}^{8}\eta_a^2=1.
\]
The corresponding conjugate-pair measurement samples $\Xi_{\mathcal A}(a)=\eta_a^2$ and attains the doubled-state QFIM.

\subsection{Tight frames and complex operator coordinates}
\label{app:complexframes}

Orthonormality is not essential. Let $\mathcal F=\{(A_a,w_a)\}_a$, with $A_a=A_a^\dagger$ and $w_a>0$, satisfy the tight Hilbert--Schmidt resolution
\begin{equation}
\sum_a
w_a
\vecop{A_a}
\langle\!\langle A_a|
=
I_{\rm HS}.
\label{eq:HermitianTightFrame}
\end{equation}
The corresponding rank-one effects are $E_a=w_a\vecop{A_a}\langle\!\langle A_a|$, and for a pure conjugate pair the outcome probabilities are
\[
\Xi_{\mathcal F}(a)
=
w_a[\Tr(A_a\rho)]^2.
\]
The tight-frame resolution gives
\begin{align*}
F_{\mu\nu}[\Xi_{\mathcal F}]
&=4\sum_aw_a\Tr(A_a\partial_\mu\rho)\Tr(A_a\partial_\nu\rho)\\
&=4\Tr[(\partial_\mu\rho)(\partial_\nu\rho)]=2F^{\rm Q}_{\mu\nu}[\rho].
\end{align*}
Thus overcompleteness leaves the pure-state Fisher geometry unchanged whenever the operator frame remains Hermitian and tight.

For comparison, let $\mathcal B=\{B_a\}_{a=1}^{d^2}$ be an orthonormal operator basis that need not be Hermitian,
\[
\Tr(B_a^\dagger B_b)=\delta_{ab},
\qquad
\zeta_a=\Tr(B_a^\dagger\rho)=u_a e^{\ii\varphi_a}.
\]
The vectorized-basis measurement produces
\[
\Gamma_{\mathcal B}(a)=|\zeta_a|^2=u_a^2.
\]
Because $\vecop{\rho}$ is normalized for a pure state and $\Tr(\rho\,\partial_\mu\rho)=0$, completeness gives
\[
2F^{\rm Q}_{\mu\nu}[\rho]
=4\sum_a\left[(\partial_\mu u_a)(\partial_\nu u_a)+u_a^2(\partial_\mu\varphi_a)(\partial_\nu\varphi_a)
\right].
\]
On the other hand,
\[
F_{\mu\nu}[\Gamma_{\mathcal B}]=4\sum_a(\partial_\mu u_a)(\partial_\nu u_a),
\]
and therefore
\[
\left[2F^{\rm Q}[\rho]-F[\Gamma_{\mathcal B}]\right]_{\mu\nu}=4\sum_a u_a^2(\partial_\mu\varphi_a)(\partial_\nu\varphi_a),
\]
which reproduces Eq.~\eqref{eq:complexdeficit}. The deficit is a positive-semidefinite Gram matrix of the coordinate phase velocities.

A generalized Weyl basis provides the canonical contrast. Its operators are unitary and its vectorized states are maximally entangled, but its coordinates $\Tr(B_a^\dagger\rho)$ are generally complex. The resulting probabilities can therefore discard parameter dependence stored in their phases even though the measurement basis is maximally entangled.

\subsection{Converse for orthogonal operator readouts}
\label{app:frameconverse}

We finally characterize when an orthogonal operator-basis readout can saturate the pure-state Fisher bound universally. Suppose $\mathcal B=\{B_a\}$ satisfies
\[
F[\Gamma_{\mathcal B}]=2F^{\rm Q}[\rho]
\]
for every pure state and every tangent direction. Equation~\eqref{eq:complexdeficit} then requires, whenever $\zeta_a\neq0$,
\[
\partial_\mu\varphi_a=0
\]
for every tangent direction, where $\zeta_a=\Tr(B_a^\dagger\rho)=u_a e^{\ii\varphi_a}$. Hence the phase of
\[
\zeta_a(\psi)=\langle\psi|B_a^\dagger|\psi\rangle
\]
is locally constant wherever it is nonzero.

Consider the numerical range
\[
W(B_a^\dagger)=\left\{\langle\psi|B_a^\dagger|\psi\rangle:
\|\psi\|=1\right\}.
\]
It is convex. If it contained two nonzero values that were not collinear in the complex plane, the segment joining them would contain points with continuously varying phase, contradicting the phase-constancy condition. Therefore $W(B_a^\dagger)$ lies on a line through the origin. Multiplication by a fixed phase $e^{-\ii\chi_a}$ rotates this line onto the real axis.

Define $\widetilde B_a=e^{-\ii\chi_a}B_a$. Then $\langle\psi|\widetilde B_a^\dagger|\psi\rangle$ is real for every $|\psi\rangle$. By the polarization identity, $\widetilde B_a$ is Hermitian. Hence
\[
B_a=e^{\ii\chi_a}\widetilde B_a,\qquad
\widetilde B_a=\widetilde B_a^\dagger.
\]
Conversely, operators that are Hermitian up to fixed phase factors have state-independent coordinate phases and eliminate the deficit in Eq.~\eqref{eq:complexdeficit}. Thus, within orthogonal operator-basis measurements, universal pure-state saturation is equivalent to Hermiticity up to fixed phases.

\section{Mixed-state geometry and the Bell-information gap}
\label{app:mixedstates}

This appendix collects the technical results underlying
Sec.~\ref{sec:mixeddistributions}. We first relate the three mixed-state
Pauli distributions, including the distinction between exact algebraic
reconstruction and statistical equivalence. We then derive the
Hilbert--Schmidt geometry of the normalized squared coordinates, prove the
exact Bell-information gap and its equality conditions, extend the
projection identity beyond Pauli operators, and give explicit mixed-state
examples.

\subsection{Mixed-state Pauli dictionaries and statistical reconstruction}
\label{app:mixeddictionary}

Expand the generally mixed state as
\[
\rho
=
\frac1d
\sum_{Q\in\mathcal P_N}
a_Q Q,
\qquad
a_Q=\Tr(Q\rho)\in\mathbb R,
\]
and define $m_Q=a_Q^2$. Then
\[
\sum_Qm_Q=dR,
\qquad
R=\Tr(\rho^2).
\]

For Pauli strings define the commutation and transpose signs by
\[
PQP=c_{PQ}Q,
\qquad
Q^\top=\tau_QQ,
\]
where $c_{PQ}=\pm1$ and $\tau_Q=(-1)^{n_Y(Q)}$. Let
$(C_N)_{PQ}=c_{PQ}$ and $D_N=\operatorname{diag}(\tau_Q)$. In the one-qubit
ordering $(I,Z,X,Y)$,
\[
C_1=
\begin{pmatrix}
1&1&1&1\\
1&1&-1&-1\\
1&-1&1&-1\\
1&-1&-1&1
\end{pmatrix},
\qquad
D_1=
\operatorname{diag}(1,1,1,-1).
\]
Tensor-product structure gives
$C_N=C_1^{\otimes N}$, $D_N=D_1^{\otimes N}$, and $C_N^2=d^2I$.

The conjugate-copy and identical-copy Bell distributions satisfy
\[
q^{\rm conj}
=
\frac1{d^2}C_Nm,
\qquad
q^{\rm id}
=
\frac1{d^2}C_ND_Nm.
\]
Since $C_N^2=d^2I$ and $D_N^2=I$,
\[
m=D_NC_Nq^{\rm id}.
\]
Hence
\[
q^{\rm conj}
=
\frac1{d^2}
C_ND_NC_Nq^{\rm id},
\qquad
\Xi_\rho^{\HS}
=
\frac1{dR}
D_NC_Nq^{\rm id}.
\]
The matrices entering these relations contain negative entries and are not
stochastic channels. Thus ideal probability vectors are algebraically
reconstructible without the corresponding measurements being statistically
equivalent.

The relevant pure and real limits follow from two fixed-point conditions.
Purity implies \(\frac1dC_Nm=m\), whereas reality in the computational basis implies \(D_Nm=m\). These relations give the four cases summarized in
Table~\ref{tab:mixeddistributions}.

\subsubsection{Finite-sample reconstruction is not direct sampling}

Define the signed map
\begin{equation}
\mathsf M
=
\frac1{d^2}C_ND_NC_N,
\qquad
q^{\rm conj}
=
\mathsf M q^{\rm id}.
\label{eq:signedBellMap}
\end{equation}
It is invertible; in fact $\mathsf M^2=I$. Suppose $N_{\rm s}$ independent
identical-copy Bell shots produce empirical frequencies
$\widehat q^{\,\rm id}$. Their multinomial covariance is
\[
\operatorname{Cov}
(\widehat q^{\,\rm id})
=
\frac1{N_{\rm s}}
\left[
D_{\rm id}
-
q^{\rm id}(q^{\rm id})^{\mathsf T}
\right],
\qquad
D_{\rm id}=\operatorname{diag}(q^{\rm id}).
\]
The reconstructed estimator
$\widehat q^{\,\rm conj}=\mathsf M\widehat q^{\,\rm id}$ is unbiased, but
\begin{equation}
\begin{aligned}
\operatorname{Cov}
(\widehat q^{\,\rm conj})
&=
\frac1{N_{\rm s}}
\mathsf M
\left[
D_{\rm id}
-
q^{\rm id}(q^{\rm id})^{\mathsf T}
\right]
\mathsf M^{\mathsf T}.
\end{aligned}
\label{eq:mappedBellCovariance}
\end{equation}
This generally differs from the covariance of direct conjugate-copy
sampling,
\[
\frac1{N_{\rm s}}
\left[
D_{\rm conj}
-
q^{\rm conj}(q^{\rm conj})^{\mathsf T}
\right],
\qquad
D_{\rm conj}=\operatorname{diag}(q^{\rm conj}).
\]

Thus $\mathsf M$ reconstructs the mean probability vector but not its
sampling law. Because the map is signed, finite-sample values of
$\widehat q^{\,\rm conj}$ may even leave the probability simplex, although
their expectation is the physical vector $q^{\rm conj}$.

More fundamentally, the transformation is one-to-one, so retaining the
complete transformed statistic cannot change the Fisher information
contained in the original count data. The likelihood after the
transformation is simply the identical-copy likelihood written in different
coordinates. It therefore carries $F[q^{\rm id}]$, not
$F[q^{\rm conj}]$. Assigning the latter would amount to replacing the actual
covariance in Eq.~\eqref{eq:mappedBellCovariance} by the multinomial
covariance of an experiment that was not performed.

For pure complex states this distinction is especially sharp:
$q^{\rm conj}=\Xi_\rho$ can be reconstructed exactly from ideal
identical-copy probabilities even though the identical-copy measurement
does not generally attain $2F^{\rm Q}[\rho]$. When $\rho=\rho^*$, by
contrast, $q^{\rm id}=q^{\rm conj}$ and the two physical experiments
coincide. As $N_{\rm s}\rightarrow\infty$, the covariance in
Eq.~\eqref{eq:mappedBellCovariance} vanishes as $N_{\rm s}^{-1}$ and the
reconstructed vector converges to the exact conjugate distribution, but the
asymptotic information per input sample remains that of the original
identical-copy experiment.

\subsubsection{Hilbert--Schmidt geometry of the normalized coordinates}

Define $\widetilde\rho=\rho/\sqrt R$. Parseval's identity gives
\[
F_{\mu\nu}[\Xi_\rho^{\HS}]
=
4\Tr\!\left[
(\partial_\mu\widetilde\rho)
(\partial_\nu\widetilde\rho)
\right].
\]
Expanding the derivative yields
\begin{equation}
\begin{aligned}
F_{\mu\nu}[\Xi_\rho^{\HS}]
&=
\frac4R
\Tr\!\left[
(\partial_\mu\rho)
(\partial_\nu\rho)
\right]
\\
&\quad
-
\frac{
(\partial_\mu R)(\partial_\nu R)
}{R^2}.
\end{aligned}
\label{eq:mixedHSmetric}
\end{equation}
This is the Hilbert--Schmidt geometry of the normalized density operator,
rather than the SLD/Bures QFIM. Neither metric generally dominates the
other in Loewner order.

For a one-parameter regular spectral branch, write
$\rho=\sum_i\lambda_i|i\rangle\langle i|$ and
$R=\sum_i\lambda_i^2$. The metric separates as
\[
F[\Xi_\rho^{\HS}]
=
F_{\rm pop}^{\HS}
+
F_{\rm coh}^{\HS},
\]
with
\[
F_{\rm pop}^{\HS}
=
4\left[
\frac{\sum_i\dot\lambda_i^2}{R}
-
\frac{
\left(\sum_i\lambda_i\dot\lambda_i\right)^2
}{R^2}
\right],
\]
and
\[
F_{\rm coh}^{\HS}
=
\frac8R
\sum_{i<j}
(\lambda_i-\lambda_j)^2
|\langle i|\dot j\rangle|^2.
\]
With the R\'enyi-2 escort weights $w_i=\lambda_i^2/R$, the population
contribution becomes
\[
F_{\rm pop}^{\HS}
=
4\,\operatorname{Var}_{w}
\!\left(
\partial_t\ln\lambda_i
\right).
\]
The corresponding coherent part of the SLD QFI is
\[
F_{\rm coh}^{\rm Q}[\rho]
=
4
\sum_{i<j}
\frac{(\lambda_i-\lambda_j)^2}
{\lambda_i+\lambda_j}
|\langle i|\dot j\rangle|^2.
\]
Thus each pair contribution to $F_{\rm coh}^{\HS}$ equals the corresponding
SLD-QFI contribution multiplied by $2(\lambda_i+\lambda_j)/R$. This makes
explicit why the normalized squared-coordinate geometry agrees with
$2F^{\rm Q}$ on the rank-one boundary but defines a different metric in the
mixed-state interior.

\subsection{Exact Bell-information gap and equality conditions}
\label{app:mixedproof}

Let $\Omega=\rho\otimes\rho^*$ and define the SLD by
$2\partial_\mu\rho=\rho L_\mu+L_\mu\rho$. For each Pauli sector set
\[
\mathcal R_P
=
\sqrt\rho\,P\sqrt\rho,
\qquad
\mathcal T_{\mu,P}
=
\sqrt\rho\,\{L_\mu,P\}\sqrt\rho.
\]
Then
\[
q^{\rm conj}(P)
=
\frac1d\Tr(\mathcal R_P^2),
\qquad
\partial_\mu q^{\rm conj}(P)
=
\frac1d
\Tr(
\mathcal R_P\mathcal T_{\mu,P}
).
\]
Therefore
\[
\begin{aligned}
[F_{\rm Bell}]_{\mu\nu}
=
\frac1d
\sum_P
\frac{
\Tr(\mathcal R_P\mathcal T_{\mu,P})
\Tr(\mathcal R_P\mathcal T_{\nu,P})
}{
\Tr(\mathcal R_P^2)
}.
\end{aligned}
\]

To resolve the doubled-state QFIM in the same Pauli sectors, expand
$\Tr(\mathcal T_{\mu,P}\mathcal T_{\nu,P})$ and use the Pauli twirl
\[
\frac1d
\sum_P
PAP
=
\Tr(A)I.
\]
Terms proportional to $\Tr(\rho L_\mu)=0$ vanish, leaving
\[
\begin{aligned}
\frac1d
\sum_P
\Tr(
\mathcal T_{\mu,P}
\mathcal T_{\nu,P}
)
&=
2F_{\mu\nu}^{\rm Q}[\rho]
\\
&=
F_{\mu\nu}^{\rm Q}[\Omega].
\end{aligned}
\]

For every sector with nonzero Bell probability define \(c_{\mu,P}
=\Tr(
\mathcal R_P\mathcal T_{\mu,P}/\Tr(\mathcal R_P^2)\), and \(\mathcal T_{\mu,P}^{\perp}=\mathcal T_{\mu,P}-c_{\mu,P}\mathcal R_P\), \(\Tr(
\mathcal R_P\mathcal T_{\mu,P}^{\perp})=0\). This decomposition is defined pointwise only on the regular support $q^{\rm conj}(P)>0$. If $q^{\rm conj}(P)=0$, then $\mathcal R_P=0$ and the ratio defining $c_{\mu,P}$ is not defined. At such a nodal sector we do not assign $c_{\mu,P}$ or $\mathcal T_{\mu,P}^{\perp}$ pointwise. Instead, Eq.~\eqref{eq:mixedgap} is understood as the continuous extension of the regular-support identity along the smooth state family, in the same sense used for vanishing Pauli probabilities in Appendix~\ref{app:zeros}. This distinction is immaterial for $\rho>0$, for which $\Tr(\mathcal R_P^2)>0$ for every Pauli string $P$, but it is required at rank-deficient and pure-state boundaries.

The Hilbert--Schmidt Pythagorean identity gives
\[
\begin{aligned}
\Tr(
\mathcal T_{\mu,P}
\mathcal T_{\nu,P}
)=
\frac{
\Tr(\mathcal R_P\mathcal T_{\mu,P})
\Tr(\mathcal R_P\mathcal T_{\nu,P})
}{
\Tr(\mathcal R_P^2)
}+
\Tr(
\mathcal T_{\mu,P}^{\perp}
\mathcal T_{\nu,P}^{\perp}
).
\end{aligned}
\]
Summing over $P$ yields Eq.~\eqref{eq:mixedgap}. Along any real
parameter-space direction $v$,
\[
\begin{aligned}
v^\mu
[
F^{\rm Q}[\Omega]-F_{\rm Bell}
]_{\mu\nu}
v^\nu
=
\frac1d
\sum_P
\|
\mathcal T_{v,P}^{\perp}
\|_{\HS}^2
\geq0.
\end{aligned}
\]
Thus the positive-semidefinite ordering follows directly from the resolved
operator geometry.

Equality along $v$ requires
\begin{equation}
\sqrt\rho\,
\{L_v,P\}
\sqrt\rho
=
c_{v,P}
\sqrt\rho\,P\sqrt\rho
\label{eq:mixedEqualityCondition}
\end{equation}
for every contributing Pauli sector. If $\rho>0$, multiplication by
$\rho^{-1/2}$ gives $\{L_v,P\}=c_{v,P}P$. Taking $P=I$ yields
$2L_v=c_{v,I}I$. Since $\Tr(\rho L_v)=0$, one obtains $L_v=0$ and therefore
$\partial_v\rho=0$. No nontrivial full-rank tangent can saturate the fixed
Bell QFI.

For a rank-one projector, $\sqrt\rho=\rho$ and
$\mathcal R_P=\Tr(P\rho)\rho$. A pure-state SLD may be chosen as
$L_\mu=2\partial_\mu\rho$, which gives
$\mathcal T_{\mu,P}=2\,\partial_\mu\Tr(P\rho)\,\rho$. Thus
$\mathcal R_P$ and $\mathcal T_{\mu,P}$ are collinear in every regular
sector, and the transverse contribution vanishes term by term.

\subsubsection{Hermitian-basis form of the mixed-state gap}

Let $\mathcal A=\{A_a\}_{a=1}^{d^2}$ be a Hermitian Hilbert--Schmidt
orthonormal basis and define
\[
\mathcal R_a
=
\sqrt\rho\,A_a\sqrt\rho,
\qquad
\mathcal T_{\mu,a}
=
\sqrt\rho\,\{L_\mu,A_a\}\sqrt\rho.
\]
The physical probabilities satisfy
\[
q_{\mathcal A}(a)
=
\Tr(\mathcal R_a^2),
\qquad
\partial_\mu q_{\mathcal A}(a)
=
\Tr(
\mathcal R_a\mathcal T_{\mu,a}
).
\]
Hilbert--Schmidt completeness gives
\[
[
F^{\rm Q}[\Omega]
]_{\mu\nu}
=
\sum_a
\Tr(
\mathcal T_{\mu,a}
\mathcal T_{\nu,a}
).
\]
Writing
$\mathcal T_{\mu,a}
=c_{\mu,a}\mathcal R_a+\mathcal T_{\mu,a}^{\perp}$
with
$\Tr(\mathcal R_a\mathcal T_{\mu,a}^{\perp})=0$
gives
\begin{equation}
\begin{aligned}
[
F^{\rm Q}[\Omega]
-
F[q_{\mathcal A}]
]_{\mu\nu}
=
\sum_a
\Tr(
\mathcal T_{\mu,a}^{\perp}
\mathcal T_{\nu,a}^{\perp}
).
\end{aligned}
\label{eq:mixedHermitianGap}
\end{equation}
Choosing $A_P=P/\sqrt d$ reduces
Eq.~\eqref{eq:mixedHermitianGap} to Eq.~\eqref{eq:mixedgap}.

\subsection{Explicit mixed-qubit example}
\label{app:mixedqubit}

For the family in Eq.~\eqref{eq:mixedQubitFamily}, the physical Bell
probabilities are
\begin{align*}
q^{\rm conj}(I)
&=
\frac{1+r^2}{4},
&
q^{\rm conj}(Y)
&=
\frac{1-r^2}{4},
\\
q^{\rm conj}(X)
&=
\frac{
1-r^2+2r^2\sin^2\vartheta
}{4},
&
q^{\rm conj}(Z)
&=
\frac{
1-r^2+2r^2\cos^2\vartheta
}{4}.
\end{align*}
The $I$ and $Y$ sectors are independent of $\vartheta$, so only $X$ and
$Z$ contribute to the Fisher information. Direct substitution gives
\[
F_{\rm Bell}(r,\vartheta)
=
\frac{
2r^4\sin^2(2\vartheta)
}{
1-r^4+r^4\sin^2(2\vartheta)
},
\]
in agreement with Eq.~\eqref{eq:mixedQubitBellFI}. Since
$F_{\vartheta\vartheta}^{\rm Q}[\rho]=r^2$, the recovered fraction is
\[
\frac{
F_{\rm Bell}
}{
2F_{\vartheta\vartheta}^{\rm Q}[\rho]
}
=
\frac{
r^2\sin^2(2\vartheta)
}{
1-r^4+r^4\sin^2(2\vartheta)
}.
\]
For $0<r<1$ this fraction can vanish while the state QFI remains finite.
At regular points it tends to unity as $r\to1$.

\subsection{Mixed spectators and other saturation cases}
\label{app:mixedexamples}

Rank-one purity of the full state is sufficient but not necessary for Bell
saturation. Consider
\[
\rho_{\boldsymbol\theta}^{SA}
=
\rho_{\boldsymbol\theta}^{S}
\otimes
\sigma^{A},
\qquad
\rho_{\boldsymbol\theta}^{S}
=
|\psi_{\boldsymbol\theta}\rangle
\langle\psi_{\boldsymbol\theta}|,
\qquad
\partial_\mu\sigma^{A}=0.
\]
For product Pauli labels $P\otimes Q$, the conjugate-copy Bell probability
factorizes as
\[
q^{\rm conj}
(P,Q;\boldsymbol\theta)
=
q_S^{\rm conj}
(P;\boldsymbol\theta)\,
r_A(Q),
\]
with $\sum_Qr_A(Q)=1$ and $\partial_\mu r_A(Q)=0$. Therefore
\[
\begin{aligned}
F_{\rm Bell}
[
\rho_{\boldsymbol\theta}^{SA}
\otimes
(\rho_{\boldsymbol\theta}^{SA})^*
]
&=
F_{\rm Bell}
[
\rho_{\boldsymbol\theta}^{S}
\otimes
(\rho_{\boldsymbol\theta}^{S})^*
]
\\
&=
2F^{\rm Q}
[
\rho_{\boldsymbol\theta}^{S}
].
\end{aligned}
\]
The parameter-independent spectator also contributes no QFI, so this equals
the doubled-state QFIM of the complete mixed system. Mixedness of an
uninformative factor therefore does not open a Bell-information gap. The
operative condition is Eq.~\eqref{eq:mixedEqualityCondition}: every
parameter-sensitive resolved tangent must remain longitudinal.

\section{Worked sensing example at equal input-copy cost}
\label{app:operationalcomparison}

This appendix gives a concrete implementation of the parameter-inference
workflow discussed in Sec.~\ref{sec:operational}. The purpose is to compare
three measurements applied to the same one-parameter family under the same
total input-copy budget: a simple fixed local readout, a one-copy SLD
measurement tailored to the working point, and the fixed two-register Bell
readout. The comparison concerns likelihood curvature and statistical
precision rather than a platform-specific experimental advantage.

Consider the pure two-qubit family
\begin{equation}
|\psi_\theta\rangle
=
e^{-\ii\theta G}|00\rangle,
\quad
G
=
\frac12
\left(
Y\otimes I+I\otimes Y
\right)
+
0.6\,X\otimes Y,
\label{eq:operationalFamily}
\end{equation}
with \(\rho_\theta=|\psi_\theta\rangle\langle\psi_\theta|\) and working point $\theta_0=0.65$.

The generator $G$ is purely imaginary in the computational basis, so
$-\ii G$ is real and the amplitudes of $|\psi_\theta\rangle$ remain real
for all $\theta$. Hence
$|\psi_\theta^*\rangle=|\psi_\theta\rangle$ for this particular family.
The physical conjugate-pair protocol can therefore be implemented here
with two identically prepared registers. This is a property of the state
family itself, not a reconstruction of conjugate-copy statistics from an
identical-copy measurement.

\subsection{Three readouts of the same parameter}

The first measurement is a simple local population readout on the first
qubit,
\[
E_0
=
|0\rangle\langle0|\otimes I,
\qquad
E_1
=
|1\rangle\langle1|\otimes I.
\]
For the family in Eq.~\eqref{eq:operationalFamily}, the corresponding
probabilities reduce to
\[
p_0^{\rm simple}(\theta)
=
\cos^2(\sqrt{0.61}\,\theta),
\qquad
p_1^{\rm simple}(\theta)
=
\sin^2(\sqrt{0.61}\,\theta),
\]
and therefore
\[
F_{\rm simple}
=
\sum_{x=0,1}
\frac{
[\partial_\theta p_x^{\rm simple}(\theta)]^2
}{
p_x^{\rm simple}(\theta)
}
=
2.44
\]
at every regular point of this trajectory.

The second measurement is the projective measurement in the eigenbasis of
the one-copy SLD evaluated at $\theta_0$. For a one-parameter pure-state
model this measurement locally attains the one-copy QFI. Since the family
is generated unitarily by the fixed operator $G$,
\[
F_{\rm SLD}
=
F^{\rm Q}[\rho_{\theta_0}]
=
4\,\operatorname{Var}_{|00\rangle}(G)
=
3.44.
\]
Unlike the simple readout, this measurement is locally optimal, but its
basis is constructed from the state model at the chosen working point.

The third measurement is the pairwise Bell readout on two copies of the
probe. Because this family is real,
$\rho_\theta^*=\rho_\theta$, so the two physical inputs also realize the
conjugate pair. The measured Pauli-string distribution therefore satisfies
\[
q^{\rm Bell}(P;\theta)
=
q^{\rm conj}(P;\theta)
=
\Xi_{\rho_\theta}(P),
\]
and its Fisher information per two-copy Bell shot is
\[
F_{\rm Bell}^{(2)}
=
2F^{\rm Q}[\rho_{\theta_0}]
=
6.88.
\]

Thus, at the working point,
\[
F_{\rm simple}=2.44,
\qquad
F_{\rm SLD}=3.44,
\qquad
F_{\rm Bell}^{(2)}=6.88.
\]
The Bell value is quoted per two-register shot, whereas the first two
values are quoted per one-copy shot.

\subsection{Likelihood and equal-copy resource accounting}

For any of the three measurements, finite counts $\{n_x\}$ define the
classical log-likelihood
\[
\ell(\theta)
=
\log\mathcal L(\theta)
=
\sum_x n_x\log p_x(\theta)
+
\text{const.},
\]
where $x$ denotes the outcomes of the chosen measurement. The maximum-
likelihood estimate satisfies
\[
0
=
\partial_\theta\ell(\hat\theta)
=
\sum_x
n_x
\partial_\theta
\log p_x(\hat\theta).
\]
The quantum structure enters only through the probability model
$p_x(\theta)$; once the measurement has been specified, the inference is an
ordinary classical estimation problem.

Near the true working point and in the large-sample regime, the normalized
likelihood has the local form
\begin{equation}
\frac{\mathcal L(\theta)}
{\mathcal L(\theta_0)}
\simeq
\exp\!\left[
-\frac12
\mathcal I_{\rm tot}(\theta_0)
(\theta-\theta_0)^2
\right],
\label{eq:localLikelihoodApprox}
\end{equation}
where
$\mathcal I_{\rm tot}=N_{\rm shots}F_{\rm shot}$.
The corresponding asymptotic standard deviation is
$\sigma_\theta\simeq\mathcal I_{\rm tot}^{-1/2}$.

Let $N_{\rm in}$ denote the total number of parameter-dependent input
registers. The simple and SLD measurements use one register per shot, so
they each permit $N_{\rm in}$ shots. The Bell protocol consumes two
registers per shot and therefore permits $N_{\rm in}/2$ Bell shots. The
three total Fisher informations are consequently
\[
\mathcal I_{\rm simple}
=
N_{\rm in}F_{\rm simple},
\]
\[
\mathcal I_{\rm SLD}
=
N_{\rm in}F^{\rm Q}[\rho_{\theta_0}],
\]
and
\[
\mathcal I_{\rm Bell}
=
\frac{N_{\rm in}}{2}
F_{\rm Bell}^{(2)}
=
N_{\rm in}F^{\rm Q}[\rho_{\theta_0}].
\]
Thus the locally tailored SLD measurement and the fixed Bell measurement
have the same Fisher rate per input register for this pure one-parameter
problem.

For
$N_{\rm in}=2\times10^5$,
\[
\sigma_{\rm simple}
\simeq
1.43\times10^{-3},
\qquad
\sigma_{\rm SLD}
\simeq
\sigma_{\rm Bell}
\simeq
1.21\times10^{-3}.
\]
The simple local readout therefore produces a likelihood approximately
$19\%$ broader than the QFI-limited alternatives, while the SLD and Bell
likelihoods have the same local width at equal input-copy cost.

\begin{table}[t]
\caption{\textbf{Equal-input-copy comparison.}
Readout resources and local precision at $\theta_0=0.65$ for the family in
Eq.~\eqref{eq:operationalFamily}, using
$N_{\rm in}=2\times10^5$ input registers. The Bell Fisher information is
quoted per two-register shot.}
\label{tab:operationalComparison}
\centering
\begin{tabular}{@{}lccc@{}}
\hline
Readout & Shots & $F$/shot & $\sigma_\theta$ \\ \hline
simple $Z_1$
& $2\times10^5$
& $2.44$
& $1.43\times10^{-3}$ \\
SLD at $\theta_0$
& $2\times10^5$
& $3.44$
& $1.21\times10^{-3}$ \\
Bell
& $10^5$
& $6.88$
& $1.21\times10^{-3}$ \\
\hline
\end{tabular}
\end{table}

\begin{figure}[!ht]
\centering
\includegraphics[width=\columnwidth]
{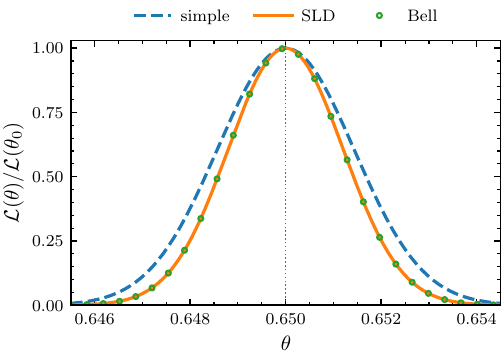}
\caption{\textbf{Likelihood comparison at equal input-copy cost.}
Local asymptotic likelihoods for the family in
Eq.~\eqref{eq:operationalFamily} at $\theta_0=0.65$, with
$N_{\rm in}=2\times10^5$. The simple local readout has a broader profile,
whereas the locally tailored SLD readout and the fixed Bell readout have
the same curvature after accounting for the two inputs consumed by each
Bell shot.}
\label{fig:operationalLikelihood}
\end{figure}

Figure~\ref{fig:operationalLikelihood} displays the corresponding local
likelihood profiles from Eq.~\eqref{eq:localLikelihoodApprox}. The curves
are not simulated finite-count records; they show the asymptotic
likelihood curvature implied by the Fisher information of each measurement.

The comparison isolates the operational distinction between the two
QFI-limited strategies. The SLD measurement is a one-copy measurement
adapted to the local working point, whereas the Bell readout is fixed and
uses two input registers per shot. In this one-parameter example they have
the same asymptotic precision at equal copy cost. The significance of the
Bell construction is therefore not a per-copy enhancement of Fisher
information, but that the quantum response is transferred to a fixed
labeled probability model that can be processed with the same classical
likelihood machinery used for an ordinary sensor.
\end{document}